\documentclass[manuscript]{aastex}

\usepackage{amsmath}
\usepackage{amssymb}
\usepackage{graphics}
\usepackage{rotating}
\usepackage{color}
\usepackage{epstopdf}
\definecolor{alizarin}{rgb}{0.82, 0.1, 0.26}
\definecolor{red}{rgb}{1.0, 0.0, 0.0}
\definecolor{green}{rgb}{0.0, 1.0, 0.0}
\definecolor{blue}{rgb}{0.0, 0.0, 1.0}

\newcommand{\ratio}{$\textrm{HCN}/\textrm{HNC}$ abundance ratio}
\newcommand{\hcn}{H$^{13}$CN}
\newcommand{\hnc}{HN$^{13}$C}
\newcommand{\aand}{$\&$ }

\newcommand{\IRASolec}{H$_{2}$}
\newcommand{\kms}{km~s$^{-1}$}
\newcommand{\ammonia}{NH$_{3}$}

\shorttitle{The HCN/HNC abundance ratio toward different evolutionary phases of massive star formation}
\shortauthors{Jin et al.}

\begin{document}

\title{The HCN/HNC abundance ratio \\ toward different evolutionary phases of massive star formation}

\author{Mihwa Jin$^1$,  Jeong-Eun Lee$^1$, and  Kee-Tae Kim$^2$} 
\affil{
       $^1$ School of Space Research, Kyung Hee University, Yongin-Si, Gyeonggi-Do 446-701, Republic of Korea; 
	    mihwajin.sf@gmail.com, jeongeun.lee@khu.ac.kr\\
       $^2$ Korea Astronomy and Space Science Institute, 776 Daedeokdae-ro, Yuseong-gu, Daejeon 305-348, Republic of Korea; ktkim@kasi.re.kr
}

\begin{abstract}

Using the \hcn\ and \hnc\ J=1--0 line observations, the abundance ratio of HCN/HNC has been estimated for different evolutionary stages of massive star formation: Infrared dark clouds (IRDCs), High-mass protostellar object (HMPOs), and Ultra-compact HII regions (UCHIIs). IRDCs were divided into `quiescent IRDC cores' and `active IRDC cores', depending on star formation activity. The HCN/HNC ratio is known to be higher at active and high temperature regions related to ongoing star formation, compared to cold and quiescent regions. Our observations toward 8 quiescent IRDC cores, 16 active IRDC cores, 23 HMPOs, and 31 UCHIIs show consistent results; the ratio is 0.97~($\pm~0.10$), 2.65~($\pm~0.88$), 4.17~($\pm~1.03$) and 8.96~($\pm~3.32$) in these respective evolutionary stages, increasing from quiescent IRDC cores to UCHIIs. The change of the HCN/HNC abundance ratio, therefore, seems directly associated with the evolutionary stages of star formation, which have different temperatures. One suggested explanation for this trend is the conversion of HNC to HCN, which occurs effectively at higher temperatures. To test the explanation, we performed a simple chemical model calculation. In order to fit the observed results, the energy barrier of the conversion must be much lower than the value provided by theoretical calculations.

\end{abstract}

\section{Introduction}

  Massive stars are believed to have a decisive effect on the physical and chemical evolution of galaxies through their energetic feedbacks, including ultraviolet (UV) radiation, stellar winds, bipolar outflows, HII regions, and eventually supernova explosions. Nevertheless, our understanding about how massive stars form is not yet well established~(see \citealt{zinnecker07} for a review). It is hard to capture the early evolutionary phases of massive stars due to their fast evolution. In addition, observational information can be easily contaminated because the objects are distant and typically form in clusters. For these reasons, observational and theoretical works regarding massive star formation are still relatively limited. Therefore, characterizing the various early phases of massive star formation in a coherent way is important to better understand massive star formation.

HCN and its geometrical isomer HNC are two fundamental and ubiquitous molecules in the dense interstellar medium. They appear to be a tracer of high-density gas, and HCN has been used as a tracer of infall motion in star forming regions~\citep{wuandevans03, sohn07}. Above all, since the \ratio\ is known to strongly depend on the kinetic temperature~\citep{goldsmith86, schilke92}, it can be a useful tool to explore the physical and chemical conditions of star-forming regions. Although both HCN and HNC are known to form in equal measure through the following dissociative recombination~\citep{mendes12},
\begin{equation}
	\textrm{HCNH}^{+} + e \rightarrow
	\begin{cases}
		\rm HCN + H
		\\
		\rm HNC + H
	\end{cases}
\end{equation}
the \ratio\ has been shown to vary by regions. For example, observational studies of the Orion Molecular Cloud  (OMC-1) showed that the abundance ratio increases from 5 at the cold edge to 80 in the warm core~\citep{schilke92}. To explain such a variation in the \ratio, many mechanisms have been suggested that the neutral--neutral reaction below, which selectively consumes HNC, is likely the main cause of the temperature dependence of the HCN/HNC abundance ratio.
\begin{equation}
\rm HNC+H \rightarrow HCN+H.
\end{equation}

\citet{schilke92} hypothesized that the above reaction is only effective at high temperature, and empirically set the rate coefficient to reproduce the variance of the HCN/HNC abundance ratio in OMC-1. In addition, \citet{talbi96} theoretically analyzed the neutral--neutral reaction using quantum--chemical calculation, and the resulting rate coefficient for the neutral--neutral reaction has been reflected in the UMIST~\citep{mcelroy13} and KIDA~\citep{harada10, harada12} chemical reaction networks with the associated conventional chemical models. \citet{talbi96} suggested that an activation energy barrier for the reaction is as high as 2000~K so that the reaction is not effective at a temperature lower than 100~K. However, this explanation does not fully account for the fluctuation of the \ratio\ in observations. For example, \citet{schilke92} and \citet{hirota98} fitted the \ratio\ for the dataset of \citet{schilke92} by modifying the rate coefficient in a semi--quantitative way. Based on the fitting, they found that the temperature dependence of the abundance ratio still holds at T $\ga$ 24K, and the rate coefficient, for which the activation energy is set to be $\thicksim$ 200~K, could well explain the observational results. Such a discrepancy between the observational works and the theoretical works still remains unresolved~\citep{graninger14}.

There have been several attempts to measure the \ratio\ toward massive star forming regions~\citep{vasyunina11, hoq13, gerner14}. However, many of those studies dealt with limited phases of massive star formation or suffered from the optical depth effect in the calculation of abundances. In this study, we determine the \ratio\ based on optically thin lines toward various evolutionary stages related to massive star formation; infrared dark clouds (IRDCs), high-mass protostellar objects (HMPOs) and ultracompact HII regions (UCHIIs).

IRDCs were identified by the MSX and Spitzer surveys as dark extinction features against the bright Galactic mid-infrared background ~\citep{egan98, simon06a, perettoandfuller09}. Since these objects contain cold ($\la$25~K) and dense ($\ga 10^5~{\rm cm}^{-3}$) cores with strong (sub)millimeter emissions (e.g., Rathborne et al. 2006), they are thought to be the ideal birthplaces of massive stars. HMPOs are luminous infrared ($L_{\rm bol} \geq 10^3~L_\odot$) point-like sources without associated radio continuum emission ~\citep{molinari96, molinari00, sridharan02, beuther02}. They have higher gas temperatures than IRDCs and mostly show outflow activities (e.g., \citet{zhang05}). Thus they are believed to be massive protostars undergoing active accretion. UCHIIs are very small (D$\lesssim$0.1 pc), dense ($n_{\rm e} \gtrsim10^4~{\rm cm}^{-3}$), and bright ($EM \gtrsim10^7~{\rm pc~cm}^{-6}$) ionized regions ~\citep{woodchurchwell89, kurtz94, segura96, ktkim01}. The objects are considered to represent the childhood of HII regions. They are the most evolved among the three phases discussed here but are still relatively early phases of massive star formation.

This paper is organized as follows. The details about source selection, observations, and used archive data are presented in \S~2. The observed spectra and the derivation of column densities of HCN and HNC and the abundance ratio between them are provided in \S~3. The analysis for the derived \ratio\ and associated chemical modeling are discussed in \S~4. The main results are summarized in \S~5.  
 
\section{Observations and archival data}

\subsection{Target selection}

After 38 IRDCs were mapped in 1.2~mm continuum emission and identified into 190 compact cores by \citet{rathborne06}, \citet{chambers09} classified them as `quiescent' prestellar cores and `active' protostellar cores by examining the star formation activity in the cores. Active IRDC cores (hereafter aIRDCc) contain both 4.5 and 24 $\micron$ infrared emissions, known as the signature of star forming activities, while quiescent IRDC cores (hereafter qIRDCc) show neither emission. Hereafter, those IRDC cores are treated in analysis as quiescent or active because each sub--sample shows these separate properties in our observational results.  
We adopted 19 qIRDCc and 35 aIRDCc from the catalog of \citet{chambers09} as our IRDC targets with following conditions: no maser sources, no nearby (32$\arcsec$) infrared sources, and well isolated from other adjacent cores. However, because of a low detection rate for the qIRDCc, it was necessary to increase the number of quiescent objects, and 13 qIRDCc were added. For the additional targets, we preferentially selected massive cores ($> 70 M_{\sun}$) because qIRDCc detected both in the \hcn\ and \hnc\ lines had their mass greater than $\sim 100 M_{\sun}$.

We selected 69 HMPOs from the catalogs of \citet{sridharan02} and \citet{molinari96} and 54 UCHIIs from the catalogs of \citet{woodchurchwell89} and \citet{kurtz94}. These selected targets have their (sub)millimeter continuum mapping data, which can be used for the \IRASolec\ column density calculation. For HMPOs, we were able to use the MAMBO 1.2~mm continuum maps from \citet{beuther02}. Consequently, our sample consisted of 67 IRDCs (32 qIRDCc and 35 aIRDCc), 69 HMPOs, and 54 UCHIIs.

 \subsection{Observations}  
 
The J=1$-$0 transitions of HCN, \hcn, HNC, and \hnc\ were observed in 2012 May, 2012 November, 2013 February, and 2014 January and March using the KVN Yonsei 21~m and the KVN Ulsan 21~m telescope \citep{ktkim11, lee11}. The lines were observed with the KVN 86~GHz SIS mixer receiver, which covers a frequency range of (85$-$95)~GHz. The rest frequencies, dipole moments, and intrinsic line strengths of observed lines are summarized in Table~\ref{observedlines}. The velocity resolution of all lines is 0.43~\kms . All observations were carried out with the position switching mode. The on-source integration time 
of the HCN and HNC lines for all sources was 10 minutes, resulting in a rms noise tempertature of $\thicksim$50~mK. For the isotopic lines, all but 13 additional qIRDCc had on-source integration time of 30 minutes.  For the 13 qIRDCc, we gave an on-source integration time of $\thicksim$1 hour because these cores appear faint in \hcn\ compared to other sub--samples. These sources were only observed in \hcn\ and \hnc\ J = 1--0. The main-beam efficiencies are 0.43 and 0.37 for the KVN Yonsei and Ulsan telescopes, respectively, and the half-power beam width (HPBW) of both telescopes is 32$\arcsec$. The system temperature ranges from 170~K to 280~K, and the pointing and focus were adjusted by observing strong SiO maser sources every one to two hours. The line intensities were calibrated by the standard chopper wheel method. Except for the 13 qIRDCc observation, we first observed all the sources in HCN and HNC J=1-0, then observed \hcn\ and \hnc\ J=1-0 lines only for the sources with ${{T_\textrm{A}}^{*}}_{,\textrm{peak}}$ $>$ $\thicksim$0.5~K in HCN and HNC J=1-0. All spectra were reduced using CLASS in the GILDAS software package.

 \subsection{Archival data}

In order to determine the fractional abundance of HCN(or HNC) relative to H$_2$, the \IRASolec\ column density must be obtained. For this, we used the MAMBO 1.2~mm continuum map data~\citep{beuther02} and the SCUBA 850 $\micron$ continuum map data~(the JCMT SCUBA legacy catalogue, \citet{james08}).

The MAMBO (37-element array) 1.2~mm data were obtained with the dual-beam OTF mapping mode and the maps typically have a resolution of 11$''$~\citep{beuther02}. The SCUBA Legacy Catalogues consist of data which exists in the JCMT archive. Because a large number of IRDCs, HMPOs, and UCHIIs in our sample have corresponding SCUBA objects counterparts, we can use these homogenous datasets throughout all three types of objects. The HPBW of JCMT is 13.5$\arcsec$ at 850 $\micron$. However, the SCUBA Legacy Catalogue data have a larger effective HPBW of 22.9$\arcsec$;~this is not only because of the effective smoothing but also because of the additional smoothing applied to reduce pixel-to-pixel noise~\citep{james08}. The flux uncertainties of both map data sets are $\thicksim$20~$\%$.

\section{Results and analysis}

\subsection{Observed spectra}

\begin{figure}
\centering
\includegraphics[width=\textwidth]{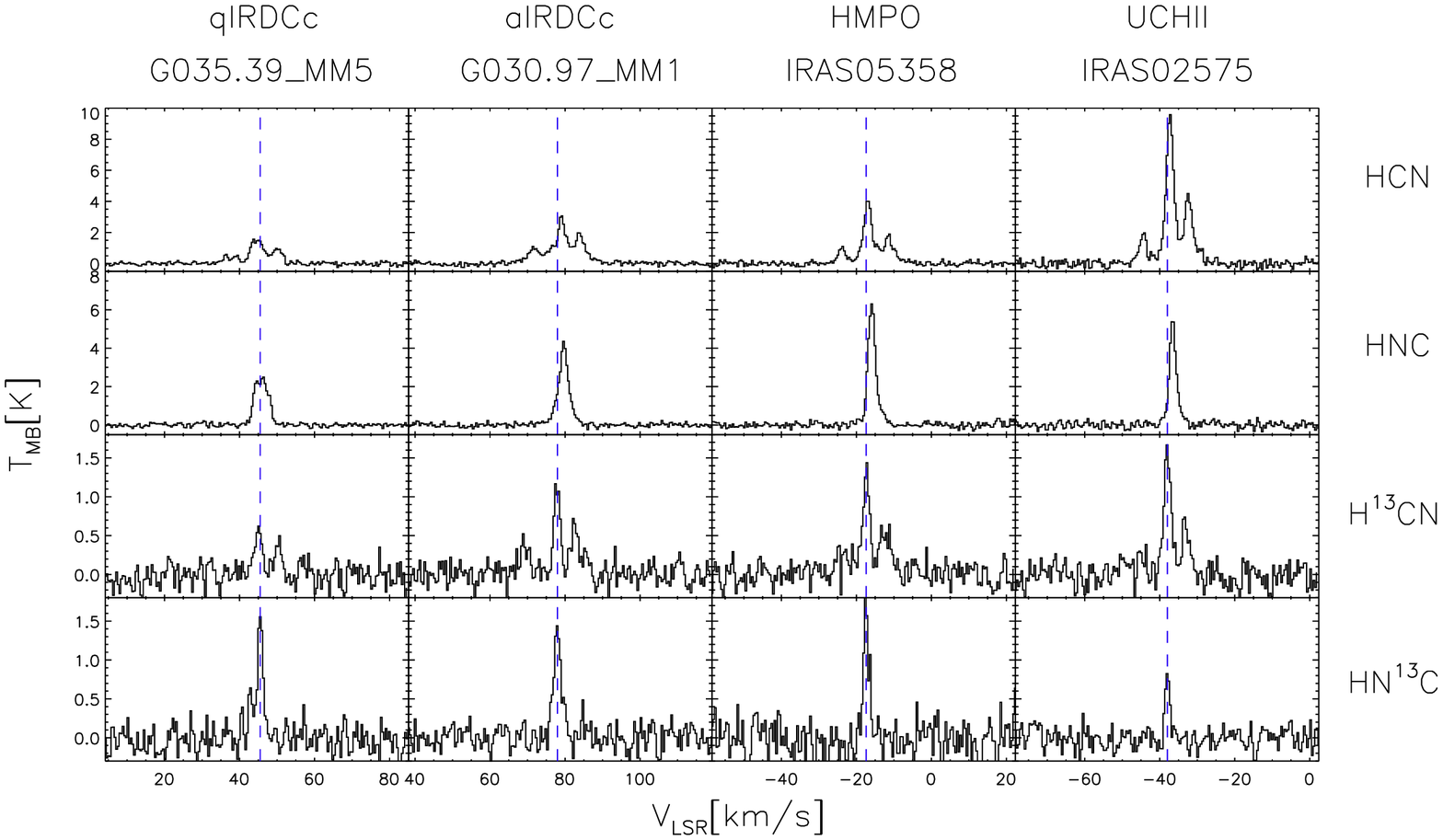}
\caption{The representative observed spectra for four sub--samples. The vertical dotted line indicates the peak velocity of the \hnc\ J=1--0 line for each source (Tables 6--9).}
\label{representplot}
\end{figure}

We detected both \hcn\ and \hnc\ lines toward 8 qIRDCc, 16 aIRDCc, 23 HMPOs and 31 UCHIIs with 3--$\sigma$ criterion. The information for these sources are listed in Tables~\ref{qirdcsourceinfo}--\ref{uchiisourceinfo}. For the \hcn\ line, the sources with the F = 2--1 hyperfine component intensity greater than the signal-to-noise ratio~(SNR) of 3 are regarded to be detected. However, because of the simple line feature of \hnc , the detection of the \hnc\ line in three sources could be confirmed by eye even though their SNR is smaller than 3. Tables~\ref{qirdcdrvdlinepara}--\ref{uchiidrvdlinepara} present the observed line parameters for qIRDCc, aIRDCc, HMPOs, and UCHIIs, respectively. All the line parameters of \hcn\ are derived from the multiple Gaussian fitting while those of \hnc\ are determined with a single Gaussian fit. Figure~\ref{representplot} exhibits the representative spectra of each sub-sample. The HCN spectra show three hyperfine components originating from the nuclear spin interaction. In optically thin and LTE conditions, the relative intensities of F = 2--1, F = 1--1, F = 1--0 hyperfine transitions of HCN should be 5~:~3~:~1, reflecting their statistical weights. However, our HCN and HNC lines seem to be optically thick and highly blended. Also, the HCN and \hcn\ lines show an anomalous hyperfine structure from what would be expected under the LTE condition~\citep{loughnane12}. The HNC and \hnc\ lines also have a hyperfine structure, but their splittings are too small ($\thicksim$0.7~\kms ;~\citet{vandertak09}) to be resolved in our sources.

Many studies have reported such HCN anomalies and have struggled to examine the underlying mechanism for those HCN anomalies~(e.g., \citealt{cernicharo84, cernicharoandguelin87, loughnane12}).
\citet{loughnane12} concluded that HCN hyperfine anomalies are common in both low-- and high--mass star-forming regions and inferred line overlap effects as the main origins of the anomalies. They have pointed out that the anomalies possibly lead to significant errors in estimating opacities with hyperfine structure fitting. As a result, the \hcn\ and \hnc\ lines were only chosen for quantitative measure of column density under the assumption that both lines are optically thin.

\begin{figure}
\centering
\includegraphics[width=\textwidth]{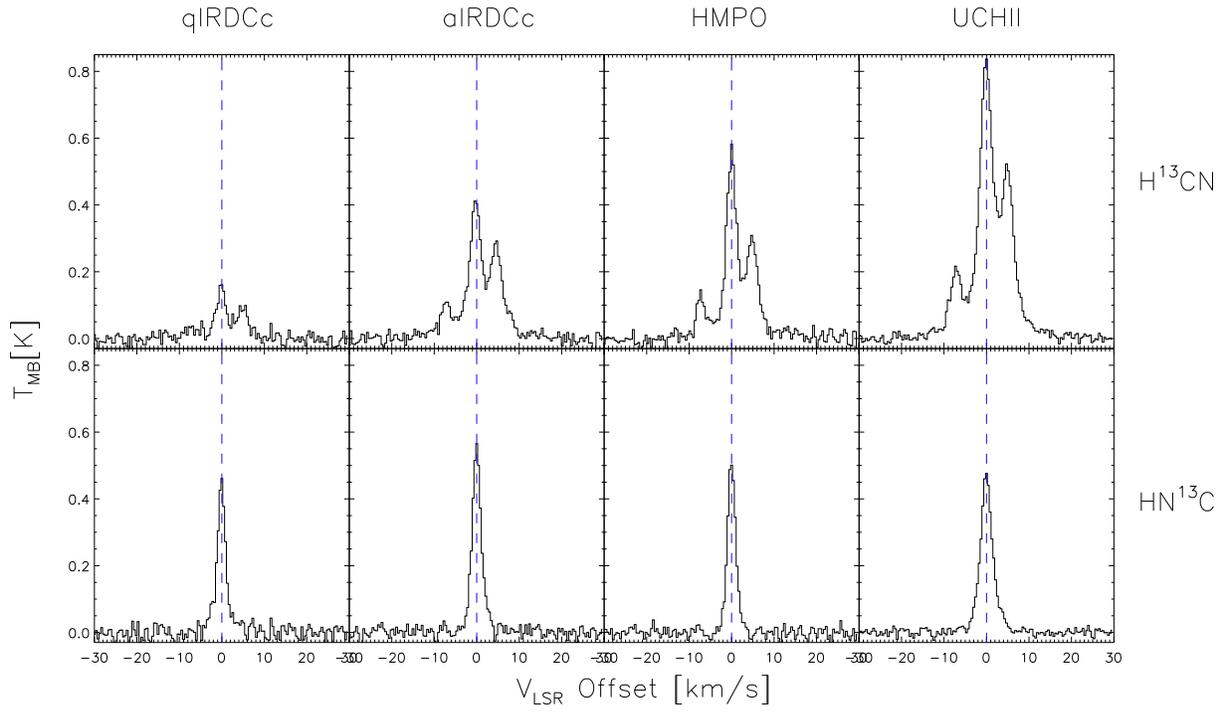}
\caption{The averaged \hcn\ and \hnc\ spectra for each sub--sample. The x-axis is relative velocity with respect to the \hnc\ peak velocity of each source, while the y--axis is beam corrected antenna temperature. }
\label{aqavrgedsptr}
\end{figure}

Figure~\ref{aqavrgedsptr} shows the averaged spectra of \hcn\ and \hnc\ for each sub-sample. The peak intensity of the averaged \hcn\ spectrum increases from qIRDCc to UCHIIs, while the \hnc\ intensity remains approximately constant, indicative of the increase of the \ratio\ with the evolutionary stage. The averaged spectra of aIRDCc and HMPOs appear to have similar peak intensities and line widths. The detailed analysis of the \ratio\ will be discussed in $\S$4.1.

\subsection{Column densities of HCN, HNC, and \IRASolec}

To determine the \ratio , we assume that the observed \hcn\ J = 1$-$0 and \hnc\ J = 1$-$0 lines arise from the same region. In order to test the reliability of the assumption, we compare the widths of two lines~\citep{sakai10};~the \hnc\ J = 1--0 line is fitted with a single Gaussian profile while the \hcn\ J = 1--0 line is fitted with its hyperfine structure to calculate the line width. In some cases, the hyperfine structure fit to the \hcn\ line was not reliable because the transitions are highly blended, and such sources were excluded by eye from the line width comparison.

\begin{figure}
\centering
\includegraphics[width=\textwidth]{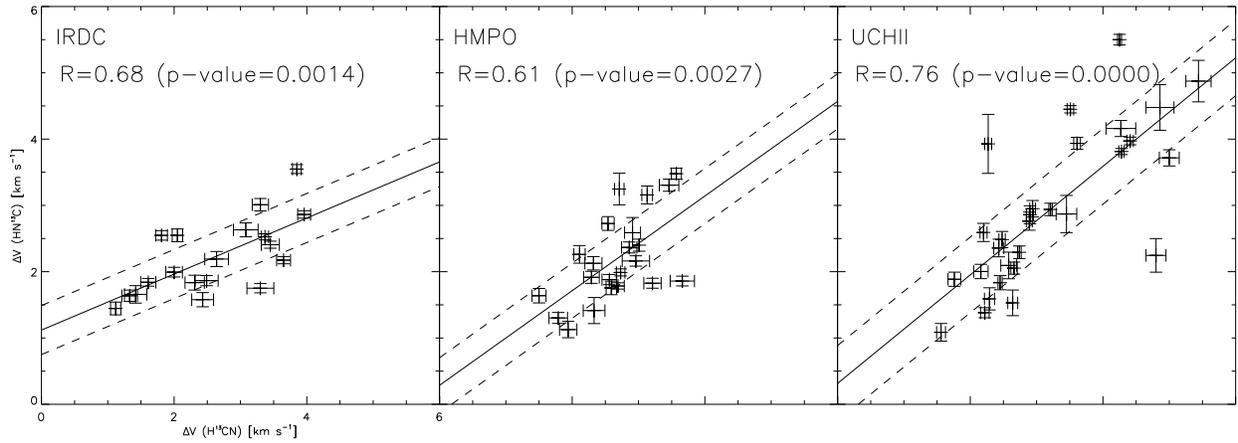}
\caption{Comparisons of the line widths between \hcn\ and \hnc . The dotted lines indicate 1--$\sigma$ for the linear regression fits. The Pearson correlation coefficient~(R) and p-value are given in the upper left corner of each box.}
\label{delvh13cnvshn13c}
\end{figure}

Figure~\ref{delvh13cnvshn13c} shows the derived line width of \hnc\ J = 1$-$0 against that of \hcn\ J = 1$-$0. All evolutionary phases show a correlation between two line widths in the confidence level above 99~$\%$. These correlations indicate that both lines arise from the same region within a clump. In this figure, we plot qIRDCc and aIRDCc together because the number of qIRDCc sources that can be used for this correlation is too small to provide a reasonable test. Above all, according to various mapping surveys of massive star forming regions including qIRDCc, the HCN and HNC lines show similar distributions in many targets~\citep{bergin97, hoq13, miettinen14, rathborne14}. This consistent spatial distribution of two line emission supports that our assumption is reasonable and thus, the \ratio\ can be used to study the physical properties of each source.

We also assume that the observed \hcn\ J = 1$-$0 and \hnc\ J = 1$-$0 lines are optically thin. In order to test the reliability of the assumption, the optical depth of these lines should be obtained. However, as mentioned in \S 3.1, opacities derived from the hyperfine structure fitting are not reliable due to the anomalous \hcn\ hyperfine structure. In addition, the hyperfine structure of \hnc\ J=1--0 is not resolved in our sources. As a result, we derived the optical depth and the excitation temperature by applying the standard LTE approximation for simplicity.

If the optically thick main isotopic line is observed, the optical depth of the \hcn\ line can be derived from
\begin{equation}
  {\tau_\textrm{thin} = - \ln\left[1-\frac{T_\textrm{thin}}{[T_\textrm{ex}-J_{\nu}(T_\textrm{bg})]}\right]}
\end{equation}
where ${T_\textrm{thin}}$ is the beam corrected brightness temperature, $T_\textrm{ex}$ is the excitation temperature derived from the optically thick main isotopic line, and $T_\textrm{bg}$ is the background radiation temperature. Since the HCN J = 1--0 emission is optically thick, we can calculate an excitation temperature according to
\begin{equation}
  {T_\textrm{ex} = \frac{h \nu_\textrm{u}}{k}\left[\ln\left(1+\frac{(h\nu_\textrm{u}/k)}{T_\textrm{thick}+J_{\nu}(T_\textrm{bg})}  \right) \right]^{-1}}
\end{equation}
where ${T_\textrm{thick}}$ is the beam corrected brightness temperature of the HCN line~\citep{purcell06}. The same calculation has been applied to the HNC and \hnc\ lines. The derived values for each source are listed in Tables~\ref{qirdcdrvdlinepara} --~\ref{uchiidrvdlinepara}. Also, the mean values \textbf{and standard deviations} of ${\tau_\textrm{thin}}$ and $ T_\textrm{ex} $ derived from individual evolutionary stages are summarized in Table~\ref{meanTauandTex}. According to the results, the $\tau_\textrm{thin}$ are as low as $\thicksim 0.1$, and thus the optically thin assumption for both isotopologues seems to be reasonable.

The derived $T_\textrm{ex}$ is considerably lower than the temperature quoted in previous work. Such low $ T_\textrm{ex} $ may suggest that the rotational levels of these molecules are excited sub--thermally. For the total column density of \hcn\ or \hnc , we adopt these excitation temperatures derived from the optically thick HCN or HNC lines as our standard; all level populations of a given molecule are determined by one excitation temperature although these molecules are sub-thermally excited. The detailed non-LTE analysis is beyond the scope of this work. However, we note that our method (i.e. an identical excitation temperature for all level populations) would overestimate column densities only by 10 $\thicksim$ 30 percent~\citep{harjunpaa04} compared to the non-LTE calculation.

The column density for those optically thin lines with one excitation temperature for all level populations can be calculated by the following equation ~\citep{lee03, mangumandshirley15}.
\begin{equation}
{N_\textrm{thin}}=\frac{{3k}{Q_\textrm{rot}}}{{8{\pi}^3}{\nu}{{\mu}^{2}}J}exp\left[\frac{E_\textrm{J}}{kT_\textrm{ex}}\right]\int{T_\textrm{R}}dv
\end{equation}
\noindent
Here $E_\textrm{J}=hBJ(J+1)$, B is the molecular rotation constant, $T_\textrm{ex}$ is the excitation temperature, $\mu$ is the electric dipole moment, and Q$_\textrm{rot}$ is the partition function. In this calculation, the beam filling factor is assumed to be unity because, according to the SCUBA legacy catalogues, all our sources that have their corresponding SCUBA data have effective radii larger than the half of our HPBW. The partition functions are calculated up to J = 10~($E_\textrm{J=10}=233.9~K$). As described above, $T_\textrm{ex}$ is derived from the main isotopic line by the equation (4). For the sources where the main isotopic lines were not observed, we adopt the mean excitation temperature for the evolutionary phase that the sources belong to.

We convert the column densities of \hcn\ and \hnc\ into those of HCN and HNC using the following isotope ratio of carbon where $D_\textrm{GC}$ is the galactocentric radius ~\citep{wilsonrood94}.
\begin{equation}
 ^{12}\textrm{C}/^{13}\textrm{C}=(7.5 \pm 1.9)D_\textrm{GC}+(7.6 \pm 12.9)
\end{equation}
To get the abundances of HCN and HNC, we also calculate the column density of molecular hydrogen using the (sub)millimeter continuum data. Since the majority of our HMPO sources were chosen from MAMBO 1.2~mm continuum survey sources, almost all of our HMPOs have corresponding MAMBO continuum maps. In addition, SCUBA continuum sources corresponding to our 4 qIRDCc, 11 aIRDCc, 18 HMPOs, and 23 UCHIIs are found. We tested 12 HMPO sources, which have both SCUBA and MAMBO maps and found that the H$_{2}$ column densities derived from different datasets are consistent with each other.

Prior to calculating the column density with the peak flux, the dust continuum maps are convolved with the KVN beam (32~$''$). Because the dust continuum emission at 850~$\micron$ and 1.2~mm is optically thin, the column density of \IRASolec\ can be calculated by
\begin{equation}
{N_\textrm{gas}}=\frac{F_{\nu}}{B_{\nu}(T_\textrm{D})\Omega \kappa_{\nu} \mu m_\textrm{H}}
\end{equation}
where $T_\textrm{D}$ is dust temperature, $\Omega$ is beam solid angle in Sr, $\mu$ is the molecular weight of the interstellar medium, and $m_\textrm{H}$ is the weight of one hydrogen atom~\citep{lee03}. $\kappa_{\nu}$ is the absorption coefficient at a given wavelength. We adopted the dust opacity $\kappa = 0.0102 ~cm^{2}g^{-1}$ at 1.2~mm~\citep{kauffmann08} and $\kappa = 0.018~cm^{2}g^{-1}$ at 850~$\micron$~\citep{lee03}. The dust temperatures are assumed to be 24~K, 35~K, 50~K and 100~K for qIRDCc, aIRDCc, HMPOs, and UCHIIs, respectively, based on the previous studies. Dust temperatures for qIRDCc and aIRDCc are adopted from \citet{rathborne10}. For HMPOs, the typical $T_\textrm{D}$ of $\thicksim$50~K, which was estimated from the SED fitting by \citet{sridharan02}, is assumed. For UCHIIs, we adopt the gas temperature that \citet{cesaroni92} have calculated using observed inversion transitions of ammonia from (1,1) to (5,5). The derived column densities, abundances, and abundance ratios are summarized in Tables~\ref{qirdccoldentable}--\ref{uchiicoldentable}.

\section{Discussion}

\subsection{Column densities and \ratio}

\begin{figure}
\centering
\includegraphics[]{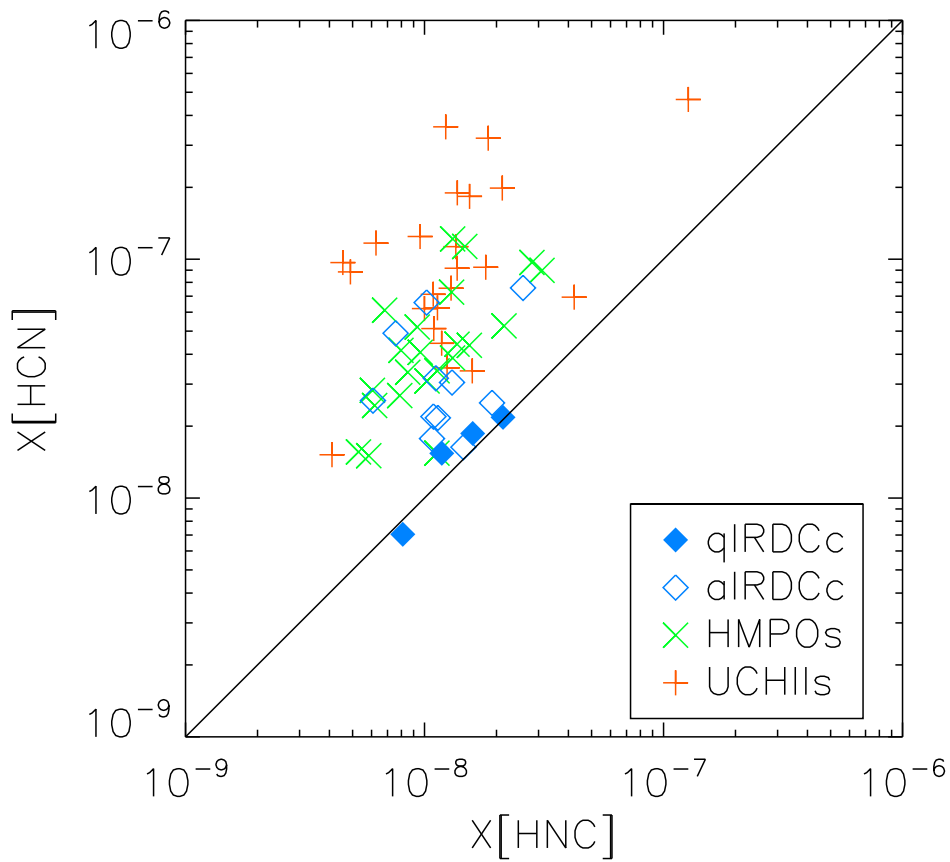}
\caption{Plot of the abundance of HCN against that of HNC. Only the sources which have their corresponding continuum map are plotted. IRDC cores, HMPOs, and UCHIIs are indicated by diamonds, times, and crosses,
respectively. Here IRDC cores are divided into quiescent ones (filled diamonds) and active ones (open diamonds). Observational uncertainty is smaller than the symbol size.}
\label{hncvshcn}
\end{figure}

 Figure~\ref{hncvshcn} compares the two fractional abundances of HCN and HNC. In the evolution from qIRDCc to UCHIIs, only the abundance of HCN increases, resulting in the increase of the \ratio . This result is consistent with \citet{sakai10}, where no systematic variation was found in the \hnc\ abundance between the MSX sources~(HMPOs) and the MSX dark sources~(IRDCs). However, this trend is opposite to the result of \citet{hirota98}, where the HCN abundance is almost constant while that of HNC decreases as the gas temperature increases. We will discuss about this discrepancy in $\S 4.2$. However, we must call attention to the fact that the abundances of species highly depend on the adopted temperatures, which have great uncertainties.

\begin{figure}
\centering
\includegraphics[width=\textwidth]{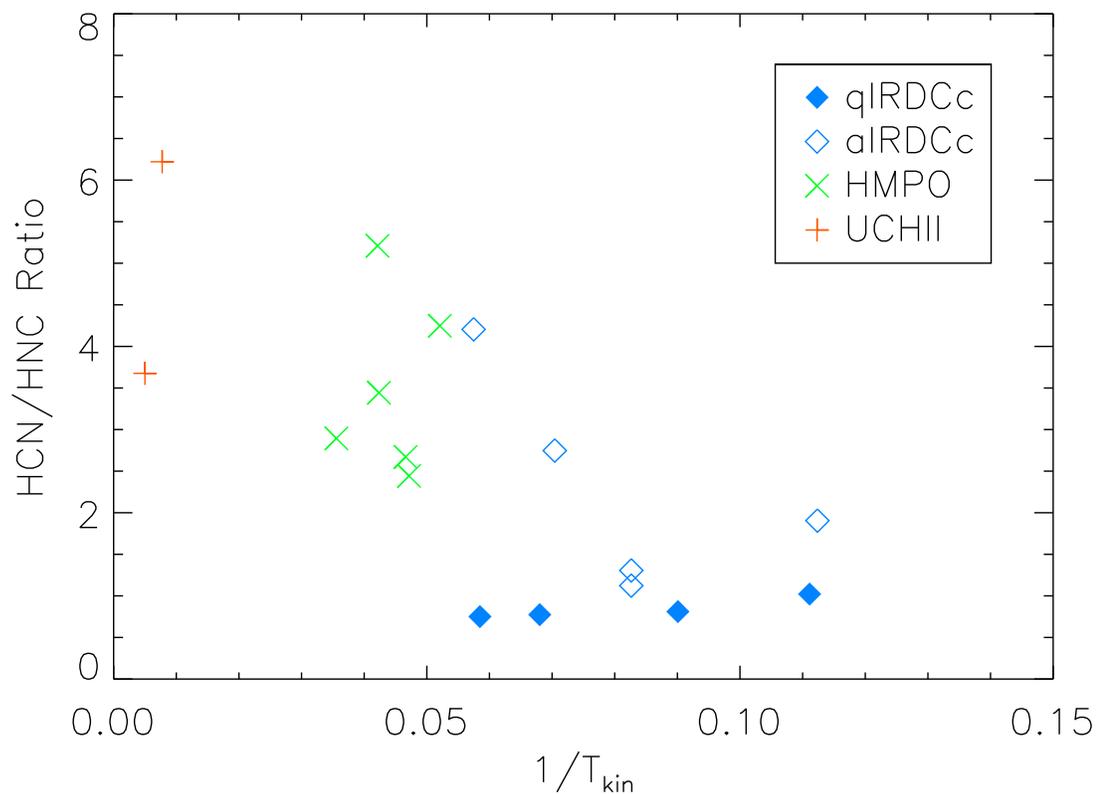}
\caption{Plot of the \ratio\ versus the kinetic temperature for our sources that have their kinetic temperature information. The kinetic temperature is adopted from \citet{chira13} for IRDCc, \citet{molinari96} for HMPO, and \citet{cesaroni92} for UCHII.}
\label{tkinvsratio}
\end{figure}

Table~\ref{drvdratio} lists the mean value and standard deviation of the \ratio\ for each evolutionary stage. The mean value gradually increases from 0.97~($\pm~0.10$), 2.65~($\pm~0.88$), 4.17~($\pm~1.03$) to 8.96~($\pm~3.32$) in the evolutionary sequence from qIRDCc to UCHIIs. This indicates an increase of temperature with evolution. Figure~\ref{tkinvsratio} shows the dependence of the \ratio\ on the kinetic temperature. The number of our sources that have the known kinetic temperature information is limited. The kinetic temperatures adopted for IRDCs and HMPOs have been calculated using the inversion transitions of ammonia (1,1) and (2,2)~\citep{chira13, molinari96} while those values for UCHIIs have been calculated using the observed inversion transitions of ammonia from (1,1) to (5,5). A similar dependence of the HCN/HNC ratio on the kinetic temperature is seen in Figure 4 of \citet{hirota98}, which claims that this dependence appears above 24 K.

\begin{figure}
\centering
\includegraphics[]{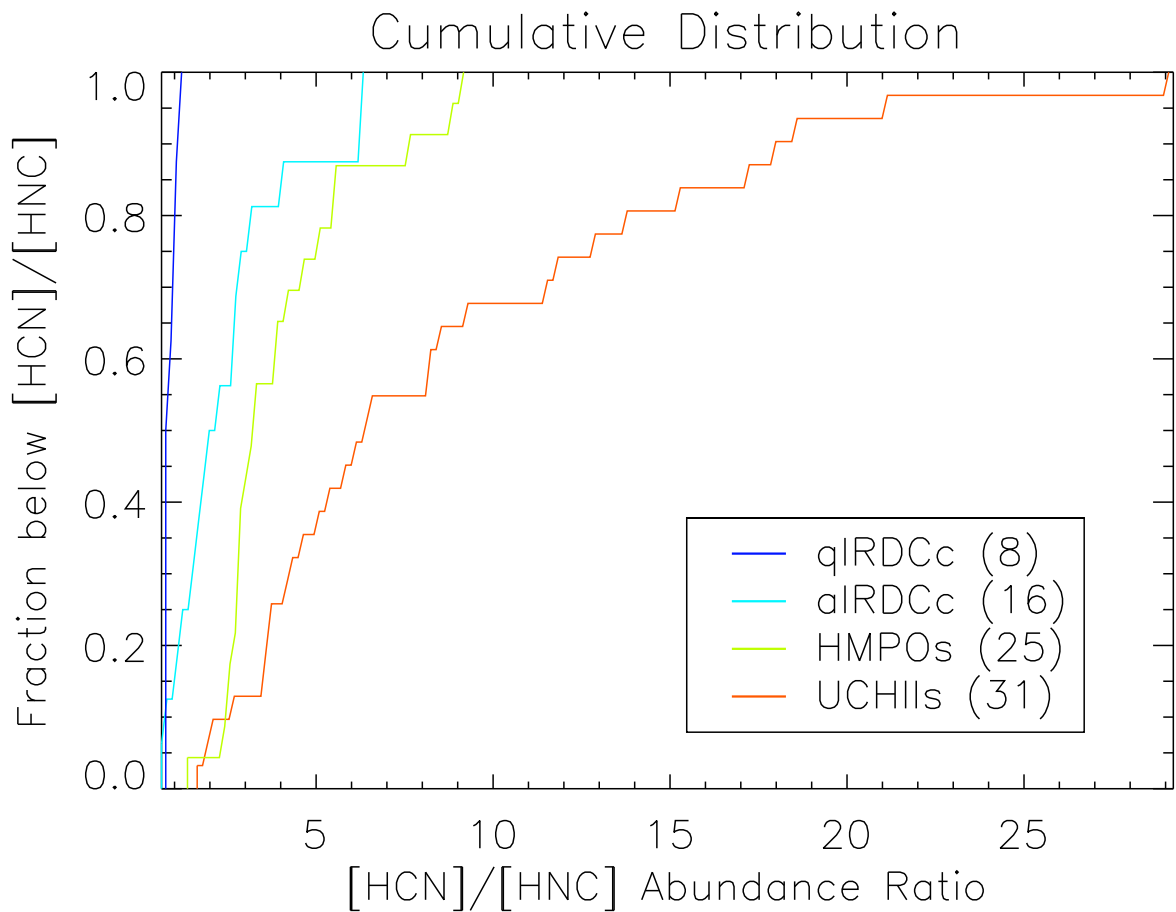}
\caption{Cumulative distribution of the \ratio . qIRDCc and UCHIIs show clearly different distributions from the other sub-samples. On the contrary, aIRDCc and HMPOs show similar distributions, implying the similar chemical conditions between the two sub-samples.}
\label{cumulratio}
\end{figure}

We also note that the mean \ratio\ of aIRDCc more coincide with the value of HMPOs rather than that of qIRDCc within the errors. The similarity between aIRDCc and HMPOs is more obvious in Figure~\ref{cumulratio}, which presents the cumulative distribution of the \ratio\ for each stage. Unlike Figure~\ref{hncvshcn}, this figure represents all the objects detected both in the \hcn\ and \hnc\ lines, because the \ratio\ only requires the column densities~(or integrated intensities) of \hcn\ and \hnc. In this figure, the distribution of the \ratio\ stretches toward a higher value as the objects evolve, but the two distributions for aIRDCc and HMPOs are alike. This result implies an analogous chemistry between the two sub-samples, which supports the suggestion that they are all in protostellar phases~\citep{chambers09, sridharan02, molinari96}. However, the mean \ratio\ of HMPO is still higher than that of aIRDCc. This can suggest that HMPOs are relatively more evolved than aIRDCc, although both sub--samples are in protostellar phases. This suggestion is consistent with the fact that HMPOs are IRAS and MSX sources~\citep{rameshandsridharan97, sridharan02}, but aIRDCc are not~\citep{simon06a}.

The separation between qIRDCc and aIRDCc, however, is not obvious in the ammonia line surveys of IRDCs. \citet{chira13} and \citet{ragan11} determined the kinetic temperatures of the IRDC cores using the (1,1) and (2,2) ammonia inversion lines and reported that there is no statistically significant difference in the derived kinetic temperature between quiescent and active cores. According to our analysis, the two types of IRDC cores have a clear difference in the \ratio , which can discriminate between gas temperatures, i.e., the aIRDCc have higher temperatures than the qIRDCc. This discrepancy between previous work and this work may be caused by the different critical densities of the observed lines. The (1,1) and (2,2) ammonia lines may trace the cooler and more extended outer envelope than the HCN J = 1--0 line because the ammonia lines have lower critical density than that of the HCN line by $\thicksim$3 orders of magnitude~\citep{berginandtafalla07, hirota98}.

There have also been several attempts to examine the chemical and physical properties of massive star forming regions using the \ratio . For example, \citet{vasyunina11} studied sub-evolutionary stages of IRDC including qIRDCc and aIRDCc based on the 3~mm molecular line data, which includes HCN, \hcn\ and HNC J = 1--0 lines. In their study, kinetic temperatures~(10~K $\thicksim$ 30~K) that come from (1,1) and (2,2) ammonia inversion line surveys were adopted as $T_\textrm{ex}$ in calculating the molecular column densities. They derived the \ratio\ of $\thicksim$1 in IRDC cores under the assumption that the HNC 1--0 line is optically thin. The result is similar to our value for qIRDCc, but there was no sign of the variation in the \ratio\ between qIRDCc and aIRDCc. \citet{sanhueza12} also studied the chemical evolution of sub-evolutionary phases in IRDCs. They reported no significant difference in the HNC abundance throughout the sub--evolutionary phases of IRDCs. We also obtained similar abundances of HNC between the two IRDC core phases, but the HCN abundances of aIRDCc were larger than that of qIRDCc, making the \ratio\ of the two IRDC cores different (see Figure~\ref{hncvshcn}).

Beyond IRDCs, \citet{hoq13} examined the HCN and HNC line data of more evolved objects associated with massive star formation. According to their results, the intensity ratio of HCN and HNC, I(HCN)/I(HNC), increased marginally from 1.07 (prestellar cores) to 1.64 (HII regions/PDRs), and the ratios were overlapped significantly between evolutionary stages. \citet{gerner14} also studied the chemical evolution from IRDC cores to UCHIIs. They showed that the estimated \ratio\ ranged from only 0.3 to 0.6. However, as seen in Figure~\ref{hncvshcn} and \ref{cumulratio}, our results are very different from these previous results; distinct separations of the \ratio\ among evolutionary stages are seen in our analyses. The main reason for the different results between our work and the previous studies will be the optical depths of the observed lines. We used the optically thin lines, \hcn\ and \hnc , while the previous studies partially or fully used the H$^{12}$CN and HN$^{12}$C lines, which are likely optically thick.

\subsection{Abundances of HCN and HNC versus adopted temperatures}

In Figure~\ref{hncvshcn}, HCN abundance varies with the evolutionary stage. However, the abundances are highly sensitive to the adopted temperatures. The higher the adopted temperature is, the higher abundance is induced. This implies that the evolutionary trend shown in the HCN abundance, increasing monotonically with evolution, possibly appears due to the adopted temperature. Excitation temperatures adopted in calculation of HCN column densities grow with evolution faster than those for HNC~(see Table~\ref{meanTauandTex}). Thus we test the HCN~(HNC) abundances versus the various sets of gas temperature.

Diverse temperature information comes both from the continuum modeling of SEDs and other molecular line observations. For example, in the case of IRDCs, \citet{chira13} and \citet{ragan11} determined $T_\textrm{kin}$ of $\thicksim$18~K using the (1,1) and (2,2) ammonia inversion lines. On the other hand, \citet{sakai08} derived $T_\textrm{rot}$ of $\thicksim$8~K using CH$_{3}$OH line. \citet{rathborne10} derived the dust temperatures of 24~K and 35~K for qIRDCc and aIRDCc, respectively, using the SED fitting method. For HMPOs, \citet{sridharan02} and \citet{molinari96} showed $T_\textrm{rot}$ of $\thicksim$23~K using the (1,1) and (2,2) ammonia inversion lines. \citet{sridharan02} derived the dust temperatures of 50~K using the modeling of SED, but \citet{beuther02} derived a much higher temperature of 100~K in their CH$_{3}$CN observations. In the case of UCHIIs, \citet{churchwell90} derived $T_\textrm{kin}$ of $\thicksim$30~K using the (1,1) and (2,2) ammonia inversion lines. \citet{cesaroni92} observed (4.4) and (5,5) inversion transitions of \ammonia\ associated with UCHII regions and concluded that the ammonia lines originate from a medium with $T_\textrm{kin} \geq$ 50~K. \citet{churchwell92} suggested higher temperatures~($\geq$ 100~K) in the molecular gas associated with UCHIIs based on the CS and CH$_{3}$CN observation data.

\begin{figure}
\centering
\includegraphics[]{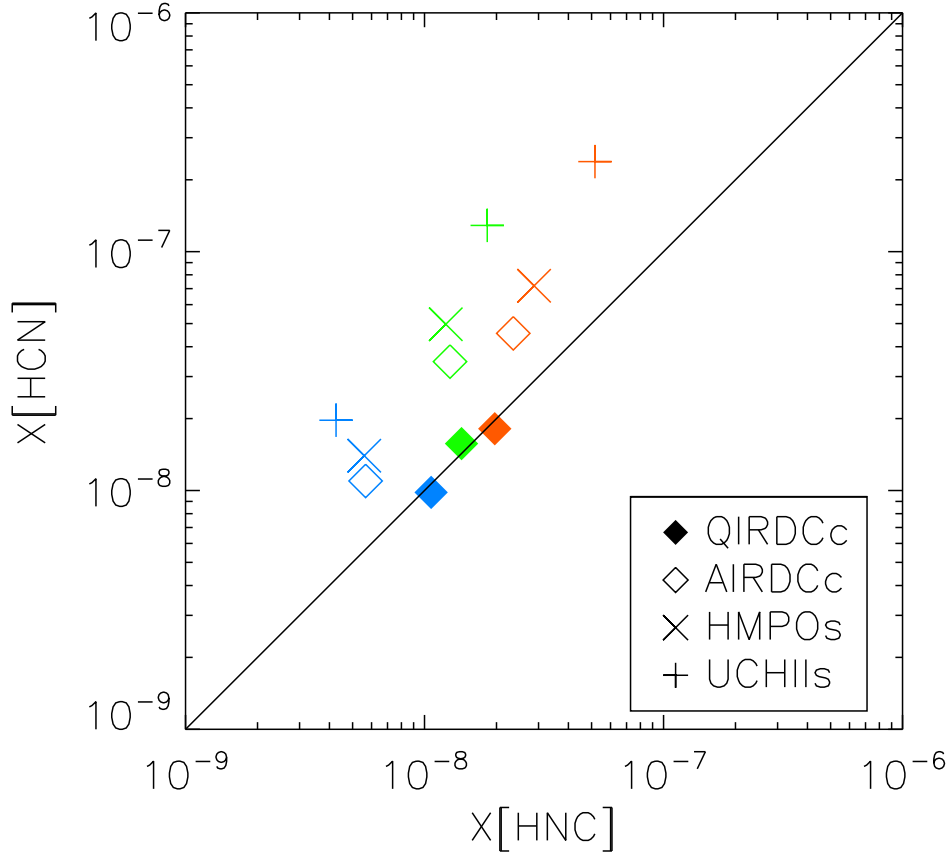}
\caption{The mean abundance of HCN against that of HNC derived from different sets of temperature. All symbols are the same as in Figure~\ref{hncvshcn}. The green symbols indicate the average abundances calculated with different excitation temperatures for \hcn\ and \hnc ~(see \S 3.2) for each evolutionary stage. However, for the red~(the highest temperature set) and blue~(the lowest temperature set) ones, we assumed the same excitation temperatures as the dust temperatures, which are fixed for each evolutionary stage. See the text for details.}
\label{temcmprsn}
\end{figure}

Figure~\ref{temcmprsn} compares the mean abundances of HCN and HNC calculated with various temperature sets. The green symbols indicate the average abundances calculated in \S 3.2 for each evolutionary stage. However, for the red and blue ones, we assumed that the excitation temperatures are the same as the gas temperatures~(i.e., the LTE condition), and adopted a fixed temperature for both molecules and for each evolutionary stage. The red symbols show the abundances derived with higher excitation temperatures, which are the same as the dust temperatures adopted in \S 3.2, while the blue ones represent the abundances calculated with a set of lower excitation temperatures. The adopted temperatures for the blue symbols are 18~K, 18~K, 23~K, 29~K for qIRDCc, aIRDCc, HMPOs, and UCHIIs, respectively. These temperatures are determined from the low transition of ammonia inversion lines~\citep{chira13, ragan11, sridharan02, molinari96, churchwell90}.

For the higher temperatures~(red symbols), both molecular abundances rapidly increase as objects evolve. However, for the lower temperatures~(blue symbols), the HCN abundance slightly increases but the HNC abundance decreases with evolution, showing a similar trend to \citet{hirota98}. This indicates that the abundance variation shown in our calculations~(Figure~\ref{hncvshcn}) could be significantly affected by the adopted temperatures.

However, even though we adopt the different sets of excitation temperature, the \ratio\ still increases from 0.83($\pm~0.06$), 2.02($\pm~0.6$) 2.65($\pm~0.5$) to 4.9($\pm~1.4$) with evolution~(blue and red symbols in Figure~\ref{temcmprsn}). \textit{This show that the increase of the \ratio\ is originated not from the adopted excitation temperatures but from the evolution of kinetic environment.}

\subsection{Chemical Modeling}

The variation of the \ratio\ with temperature has been a debated issue. The \ratio\ has been calculated in various sources, from dark clouds~\citep{hirota98}, active star--forming region~\citep{schilke92}, to external galaxies and associated large--scale outflows~\citep{aalto12}. \citet{schilke92} suggested that the neutral--neutral reaction described in Equation (2) mainly governs the transition of the \ratio , and set the activation energy of the reaction to be 200~K, which best reproduces the \ratio\ observationally determined. On the other hand, \citet{talbi96} quantum--chemically analyzed the neutral--neutral reaction and suggested that the reaction should have an activation energy as high as 2000~K, resulting in a rate coefficient of $\thicksim\ 10^{-15}$ at a temperature lower than 100~K. However, this rate cannot reproduce the observed results. The discrepancy between the rate coefficients empirically determined~(hereafter $k_\textrm{Schilke}$) and theoretically determined~(hereafter $k_\textrm{Talbi}$) has remained unresolved~\citep{graninger14}.

We tested both rate coefficients for our chemical calculations and compared them with our observations. We modeled the \ratio\ evolution using the astrochemical code, ALCHEMIC~\citep{semenov10} with the UMIST gas chemical reaction network~\citep{mcelroy13}. The only surface chemistry considered in our calculation is the \IRASolec\ formation, and its rate equation has been adopted from \citet{cazauxandtielens04, cazauxandtielens10}. The chemical evolution has been calculated adopting three continuous evolutionary stages of physical conditions: from IRDCs, through HMPOs, to UCHIIs. The physical conditions at a given evolutionary stage are assumed constant. The initial chemical abundances for the HMPO stage were adopted from the final chemical abundances for the IRDC stage, and the final chemical abundances for the stages of HMPOs were taken as the initial abundances for the UCHII stage. The initial abundances for the IRDC stage were adopted from the set with low metals~(Table~7 of \citet{gerner14}). However, the atomic abundance of N is decreased from $2.47\times10^{-5}$ to $1.00\times10^{-5}$ to achieve a better fit to the calculated abundances. The physical condition and chemical timescale adopted for each phase are listed in Table~\ref{modelcondition}. Since the chemical evolution is associated with kinetic temperature, in our chemical calculations, we assume that the gas temperatures are the same as the dust temperatures adopted in \S 3.1. In this dense environment, gas and dust are thermally well coupled. Actually we tested a variety of gas temperature in our chemical calculation. According to our test, lower gas temperatures adopted for the blue symbols in Figure~\ref{temcmprsn} cannot reproduce the observed $\textrm{HCN}/\textrm{HNC}$ abundance ratios.

The model mainly solves the following equation for 4604 gas--phase reactions.
\begin{equation}
\frac{\textrm{d}n(i)}{\textrm{d}t}=\sum_{l,m}k_{lm}n_{l}n_{m}-n_{i}\sum_{i \ne l}k_{l}
\end{equation}
where $n(i)$ is the gas--phase number density of the $i$--th species ($\rm cm^{-3}$), and $k_{lm}$ and $k_{l}$ are rates of the gas--phase reactions. For the first--order kinetics and for the second--order kinetics, the units are $\rm s^{-1}$ and $\rm cm^{3}~s^{-1}$, respectively~\citep{semenov10}. The cosmic ray~(CR) ionization rate is assumed to be $5.0\times10^{-17}$~s$^{-1}$. For simplicity, the effect by UV photons was not included, i.e., the model gas is assumed to be well shielded from either external or internal UV radiation. This assumption could be reasonable because HCN traces very dense gas; the dense inner region of prestellar cores will be shielded from the external interstellar radiation field, which is the only UV radiation source in this stage. On the other hand, in the protostellar stage, the UV photons emitted from the central protostars will be absorbed by the very dense infalling material, which prevents the high energy photons from traveling farther out. In addition, the UV photons from the early type central stars in UCHIIs are absorbed by the inner warm dusty envelope,  which converts the entire stellar luminosity to FIR radiation~\citep{churchwell02}.

\begin{figure}
\centering
\includegraphics[angle=90,origin=c, width=\textwidth]{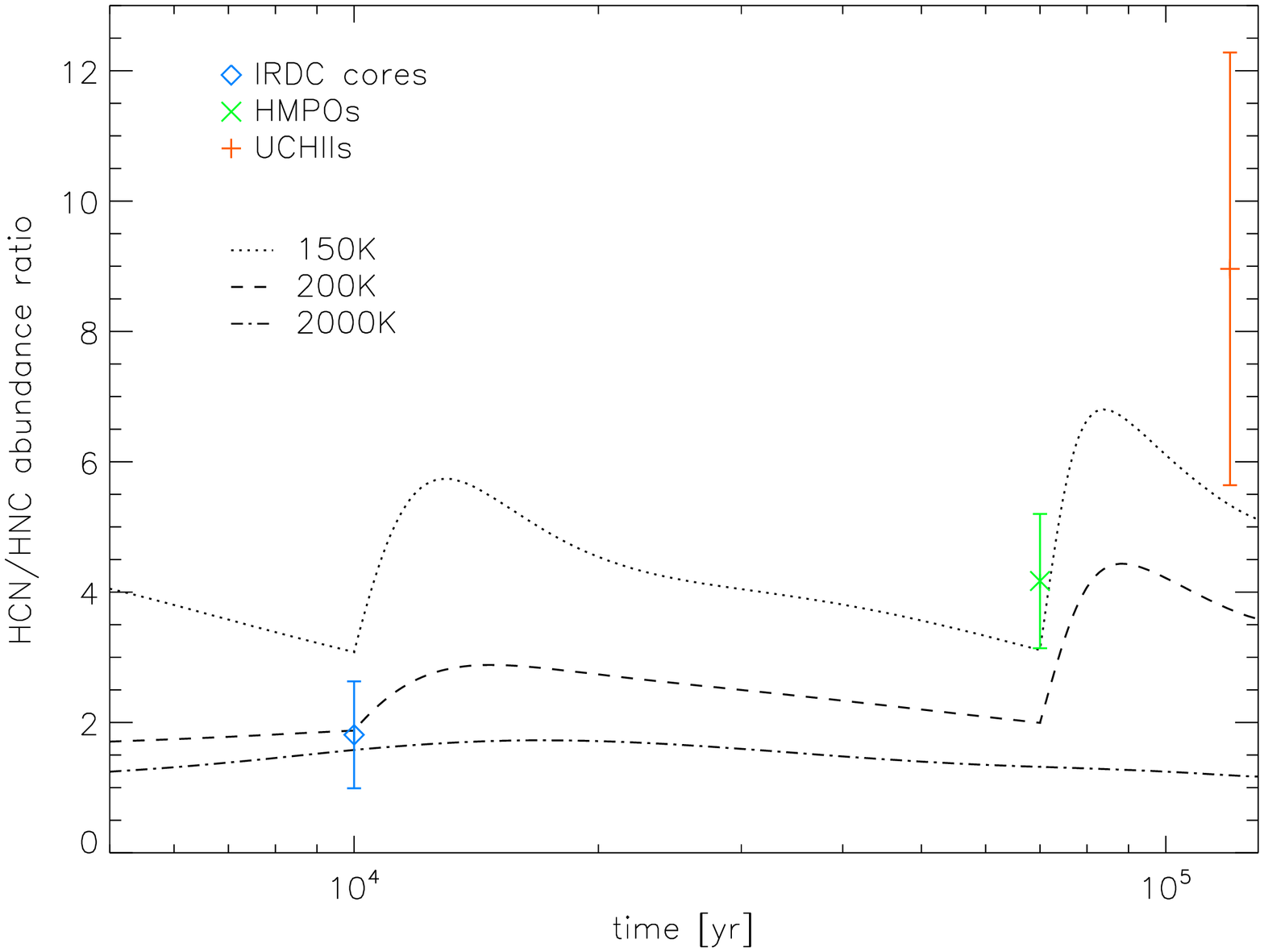}
\caption{A comparison between the \ratio\ derived from the observation and the ratios calculated from the chemical model. Dashed, dotted, and solid lines indicate the calculated ratios with activation barriers of 2000~K~\citep{talbi96}, 200~K~\citep{schilke92}, and 150~K, respectively. The observed ratios for IRDCs, HMPOs, and UCHIIs are indicated by diamonds, times, and crosses, respectively. For the each evolutionary stage, chemical timescales of \citet{gerner14} are adopted.}
\label{chemresult}
\end{figure}

Figure~\ref{chemresult} compares the observed \ratio\ with the calculated values. Here the chemical time scales are taken from \citet{gerner14}. Different line types indicate different activation energies used in the chemical model. For the observed \ratio\ of IRDCs, we adopt the mean ratio of all qIRDCc and aIRDCc since the timescale for each of those cores is not well known. The calculated \ratio\ is controlled by the activation energy for the neutral--neutral reaction, and that $k_\textrm{Schilke}$ provides results much more consistent with our observational ratios compared to $k_\textrm{Talbi}$. However, the observed ratios are better fitted by the model with an activation energy of 150~K, which results in a rate coefficient higher than $k_\textrm{Schilke}$. In this chemical calculations, we did not try to fit the exact abundances of HCN and HNC since the abundances calculated from observations are very sensitive to the adopted excitation temperature as mentioned in the previous section.

The model we adopt here is too simple to provide a precise analysis; only three discrete physical conditions have been adopted to describe the evolution of massive star formation, and the timescale for each evolutionary stage has large uncertainty. Nevertheless, our test with chemical models clearly show that (1) the \ratio\ can be a evolutionary tracer of massive star formation and (2) to fit the ratios derived from observations, the activation energy for the neutral--neutral reaction should be much smaller than the theoretical value.

\section{Summary}

To examine evolutionary effects in the chemical and physical conditions of massive star-forming regions, we carried out \hcn\ and \hnc\ line observations and analyzed the \ratio\ toward IRDC cores~(qIRDCc and aIRDCc), HMPOs, and UCHIIs, which are known as early phases of massive star formation. We detected both \hcn\ and \hnc\ lines toward 8 qIRDCc, 16 aIRDCc, 23 HMPOs, and 31 UCHIIs.

1. We found a statistically increasing tendency of the abundance ratio with the evolution of the objects, from 0.97 (qIRDCc) to 8.96 (UCHIIs). This follows the known evolutionary scheme for massive star formation well and indicates that the \ratio\ can be used to trace the evolutionary stages of massive star formation, in which temperature increases.

2. aIRDCc and HMPOs show similar mean values and distributions of the \ratio. This implies that the two types of sources have similar chemical conditions and supports the suggestion that both types of sources may be in the protostellar phase.

3. Although aIRDCc and HMPO show similar abundance ratios, the mean value of HMPOs is still higher than that of aIRDCc. This may indicate that both sub--samples are in the same protostellar phase, but HMPOs are relatively more evolved than aIRDCc. This suggestion is consistent with the fact that HMPOs are associated with MSX sources, while aIRDCc are not.

4. Our observed \ratio\ cannot be explained by the theoretically determined rate coefficients of the neutral--neutral reaction. This discrepancy between observational and theoretical works remains to be resolved. 

\section*{Acknowledgements}

We are grateful to all staff members in KVN who helped to operate the array. The KVN is a facility operated by the Korea Astronomy and Space Science Institute. This research was supported by the Basic Science Research Program through the National Research Foundation of Korea (NRF) funded by the Ministry of Education of the Korean government (grant No. NRF-2012R1A1A2044689). This work was also supported by the Korea Astronomy and Space Science Institute (KASI) grant funded by the Korea government (MEST) and the BK21 plus program through the NRF funded by the Ministry of Education of Korea.

\clearpage

\clearpage

\clearpage

\clearpage

\clearpage

\clearpage

\clearpage

\clearpage

\begin{deluxetable}{lclll}
\tablewidth{0pt}
\tabletypesize{\footnotesize}
\tablecolumns{5}
\tablecaption{Observed Lines}
\scriptsize
\tablehead{
\colhead{Molecule} 		& \colhead{Transition}	 	& \colhead{$\nu$ [MHz]}	 & \colhead{$\mu$ [D]}		& \colhead{$S_\textrm{ul}^\textrm{a}$}
}
\startdata
HCN & J=1-0, F=2--1 & 88630.42 & 2.99$^\textrm{b}$ & 3\\
 &J=1--0, F=1--1 & 88631.85 &  & 5\\
 & J=1--0, F=0--1 & 88633.94 &   & 1\\
\hcn\ & J=1-0, F=2--1 & 86338.77 & 2.99$^\textrm{b}$ & 3\\
 &J=1--0, F=1--1 & 86340.18 &  & 5\\
 & J=1--0, F=0--1 & 86342.27 &   & 1\\
HNC & J=1--0 & 90663.57 & 3.05$^\textrm{c}$  & \\ 
\hnc\ & J=1--0 & 87090.85 & 3.05$^\textrm{c}$  & \\
\enddata
\label{observedlines}
\tablecomments{$^\textrm{a}$ Intrinsic line strength. $^\textrm{b}$ \citet{bhattacharyaandgordy60}. $^\textrm{c}$ \citet{blackman76}.}
\end{deluxetable}

\begin{deluxetable}{l c c c c c c c c}
\tablewidth{0pt}
\tabletypesize{\footnotesize}
\tablecolumns{          9}  
\tablecaption{qIRDCc source Information}  
\scriptsize
\tablehead{\colhead{Source Name} & \colhead{RA} & \colhead{Dec} & \colhead{$l$} & \colhead{$b$} & \colhead{$D_\textrm{kin}$} & \colhead{$D_\textrm{GC}$} & \colhead{$R$} & \colhead{$T_\textrm{gas}$} \\
\colhead{} & \colhead{J(2000.0)}  & \colhead{J(2000.0)}  &  \colhead{[$\arcdeg$]} & \colhead{[$\arcdeg$]} & \colhead{[$kpc$]} & \colhead{[$kpc$]} & \colhead{[$\arcsec$]} & \colhead{[$K$]} }
\startdata

                G018.82$\_$MM6    & 18:26:18.4      &           -12:41:15     & 18.80    & -0.29    &  4.8    &  5.7    &   \nodata    &             \nodata  \\[-1ex]
                G028.37$\_$MM9    & 18:42:46.7      &           -04:04:08     & 28.32    &  0.07    &  5.0    &  5.3    & 57.1    &             \nodata  \\[-1ex]
               G028.37$\_$MM17   & 18:43:00.0      &           -04:01:34     & 28.39    &  0.04    &  5.0    &  5.3    & 40.7    &          $9.3$  \\[-1ex]
                G028.53$\_$MM7    & 18:44:23.7      &           -04:02:09     & 28.54    & -0.28    &  5.7    &  5.1    &   \nodata    &         $11.7$  \\[-1ex]
                G031.97$\_$MM9    & 18:49:31.6      &           -00:46:30     & 32.02    &  0.07    &  6.9    &  5.1    &   \nodata    &             \nodata  \\[-1ex]
                G033.69$\_$MM8    & 18:52:53.9      &           +00:41:16     & 33.71    & -0.01    &  7.1    &  5.0    & 27.9    &             \nodata  \\[-1ex]
                G034.77$\_$MM4    & 18:56:48.9      &           +01:23:34     & 34.78    & -0.56    &  2.9    &  6.7    &   \nodata    &         $16.3$  \\[-1ex]
                G035.39$\_$MM5    & 18:57:08.8      &           +02:08:09     & 35.48    & -0.30    &  2.9    &  6.7    &   \nodata    &         $19.7$  \\
\enddata
\tablecomments{Kinetic distances are quoted from \citet{simon06b}. The source sizes~($R$) are quoted from effective radius in the JCMT SCUBA legacy catalogue \citep{james08}. The ${T_\textrm{gas}}$ are quoted from ${T_\textrm{kin}}$ in \citet{chira13}.}
\label{qirdcsourceinfo}
\end{deluxetable}

\begin{deluxetable}{l c c c c c c c c}
\tablewidth{0pt}
\tabletypesize{\footnotesize}
\tablecolumns{          9}  
\tablecaption{aIRDCc source Information}  
\scriptsize
\tablehead{\colhead{Source Name} & \colhead{RA} & \colhead{Dec} & \colhead{$l$} & \colhead{$b$} & \colhead{$D_\textrm{kin}$} & \colhead{$D_\textrm{GC}$} & \colhead{$R$} & \colhead{$T_\textrm{gas}$} \\
\colhead{} & \colhead{J(2000.0)}  & \colhead{J(2000.0)}  &  \colhead{[$\arcdeg$]} & \colhead{[$\arcdeg$]} & \colhead{[$kpc$]} & \colhead{[$kpc$]} & \colhead{[$\arcsec$]} & \colhead{[$K$]} }
\startdata
                G019.27$\_$MM2    & 18:25:52.6      &           -12:04:48     &19.29    &  0.08    &  2.4    &  6.6    & 41.0    &          $9.2$  \\[-1ex]
                G022.35$\_$MM1    & 18:30:24.4      &           -09:10:34     &22.38    &  0.45    &  4.3    &  5.7    & 58.0    &         $15.8$  \\[-1ex]
                G023.60$\_$MM1    & 18:34:11.6      &           -08:19:06     &23.57    &  0.01    &  3.9    &  4.3    &   \nodata    &             \nodata  \\[-1ex]
                G024.60$\_$MM1    & 18:35:40.2      &           -07:18:37     &24.63    &  0.15    &  3.7    &  5.9    & 49.3    &         $20.2$  \\[-1ex]
                G028.37$\_$MM4    & 18:42:50.7      &           -04:03:15     &28.34    &  0.06    &  5.0    &  5.0    & 51.0    &         $13.0$  \\[-1ex]
                G028.37$\_$MM6    & 18:42:49.0      &           -04:02:23     &28.35    &  0.07    &  5.0    &  5.3    & 40.9    &             \nodata  \\[-1ex]
                G030.97$\_$MM1    & 18:48:21.6      &           -01:48:27     &30.97    & -0.14    &  5.1    &  5.4    & 54.8    &             \nodata  \\[-1ex]
                G031.97$\_$MM1    & 18:49:36.3      &           -00:45:45     &32.04    &  0.06    &  6.9    &  5.1    & 53.3    &             \nodata  \\
\enddata
\tablecomments{Kinetic distances are quoted from \citet{simon06b}.The source sizes~($R$) are quoted from effective radius in the JCMT SCUBA legacy catalogue \citep{james08}. The ${T_\textrm{gas}}$ are quoted from ${T_\textrm{kin}}$ in \citet{chira13}.}
\label{airdcsourceinfo}
\end{deluxetable}

\begin{deluxetable}{l c c c c c c c c c}
\tablewidth{0pt}
\tabletypesize{\footnotesize}
\tablecolumns{          10}  
\tablecaption{HMPO source Information}  
\scriptsize
\tablehead{\colhead{Source Name} & \colhead{RA} & \colhead{Dec} &  \colhead{$l$} & \colhead{$b$} & \colhead{$D_\textrm{kin,far}$} &\colhead{$D_\textrm{kin,near}$} & \colhead{$D_\textrm{GC}$} & \colhead{$R$} & \colhead{$T_\textrm{gas}$} \\
\colhead{} & \colhead{J(2000.0)}  & \colhead{J(2000.0)}  & \colhead{[$\arcdeg$]} & \colhead{[$\arcdeg$]} & \colhead{[$kpc$]} & \colhead{[$kpc$]} & \colhead{[$kpc$]} & \colhead{[$\arcsec$]} & \colhead{[$K$]} }
\startdata
                IRAS05358+3543    & 05:39:10.4      &           +35:45:19     &173.48    &  2.43    &  1.8    &  \nodata    & 21.2    & 47.3    &         $18.0^{b}$  \\[-1ex]
                IRAS05373+2349    &05:40:24.4      &           +23:50:54     &183.72    & -3.66    &  1.2    &  \nodata    &  9.8    & 42.5    &         $21.2^{a}$  \\[-1ex]
                IRAS18024-2119    &  18:05:25.4      &           -21:19:41     &8.83    & -0.03    &  0.1   &  \nodata    &  8.4    &  \nodata    &          \nodata  \\[-1ex]
                IRAS18089-1732    & 18:11:51.3      &           -17:31:28     &12.89    &  0.49    & 13.0    &  3.6    &  5.5    & 44.9    &         $19.2^{a}$  \\[-1ex]
                IRAS18102-1800    & 18:13:12.2      &           -17:59:35     &12.63    & -0.02    & 14.0    &  2.6    &  6.3    &  \nodata    &         $15.0^{b}$  \\[-1ex]
                IRAS18144-1723    & 18:17:24.4      &           -17:22:13     &13.66    & -0.60    &  4.3    &  \nodata    &  4.9    & 58.3    &         $23.6^{a}$  \\[-1ex]
                IRAS18151-1208    & 18:17:57.1      &           -12:07:22     &18.34    &  1.77    &  3.0    &  \nodata    &  6.2    &  \nodata    &         $23.7^{a}$  \\[-1ex]
                IRAS18182-1433    & 18:21:07.9      &           -14:31:53     &16.58    & -0.05    & 11.8    &  4.5    &  4.8    & 38.0    &         $20.0^{b}$  \\
\enddata
\label{hmposourceinfo}
\tablecomments{Kinetic distances are quoted from \citet{beuther02} and from \citet{molinari96}. If the distance ambiguity is resolved, only far and no near distance is noted. The source sizes~($R$) are quoted from effective radius in the JCMT SCUBA legacy catalogue \citep{james08}. The ${T_\textrm{gas}}^{a}$ are quoted from $T_\textrm{kin}$ in \citet{molinari96} and the The ${T_\textrm{gas}}^{b}$ are quoted from ${T_\textrm{rot}}$ in \citet{beuther02}.} 
\end{deluxetable}

\begin{deluxetable}{l c c c c c c c c}
\tablewidth{0pt}
\tabletypesize{\footnotesize}
\tablecolumns{          9}  
\tablecaption{UCHII source Information}  
\scriptsize
\tablehead{\colhead{Source Name} & \colhead{RA} & \colhead{Dec} & \colhead{$l$} & \colhead{$b$} & \colhead{$D_\textrm{kin}$} & \colhead{$D_\textrm{GC}$} & \colhead{$R$} & \colhead{$T_\textrm{gas}$} \\
\colhead{} & \colhead{J(2000.0)}  & \colhead{J(2000.0)}  &  \colhead{[$\arcdeg$]} & \colhead{[$\arcdeg$]} & \colhead{[$kpc$]} & \colhead{[$kpc$]} & \colhead{[$\arcsec$]} & \colhead{[$K$]} }
\startdata
                IRAS02232+6138    & 02:27:01.0      &           +61:52:14     &133.94    &  1.06    &  3.0    & 11.4    &  \nodata    &             \nodata  \\[-1ex]
                IRAS02575+6017    & 03:01:32.3      &           +60:29:12     &138.30    &  1.56    &  3.8    & 11.0    & 62.0    &             \nodata  \\[-1ex]
                IRAS03035+5819    & 03:07:25.6      &           +58:30:52     &139.91    &  0.20    &  4.2    & 11.2    & 48.0    &             \nodata  \\[-1ex]
                IRAS06053-0622    & 06:07:46.6      &           -06:22:59     &213.70    &-12.60    & 10.8    &  9.2    &  \nodata    &             \nodata  \\[-1ex]
                IRAS06084-0611    & 06:10:51.0      &           -06:11:54     &213.88    &-11.84    &  1.0    &  9.3    &  \nodata    &             \nodata  \\[-1ex]
                IRAS06056+2131    &06:08:41.0      &           +21:31:01     &189.03    &  0.78    &  0.8    &  8.6    & 63.4    &             \nodata  \\[-1ex]
                IRAS06058+2138    &06:08:54.1      &           +21:38:25     &188.95    &  0.89    &  2.2    &  9.2    &  \nodata    &             \nodata  \\[-1ex]
                IRAS06061+2151    &06:09:07.8      &           +21:50:39     &188.80    &  1.03    &  4.1    &  8.5    & 59.4    &             \nodata  \\
\enddata
\tablecomments{Kinetic distances are quoted from \citet{thompson06} and references therein. The source sizes~($R$) are quoted from effective radius in the JCMT SCUBA legacy catalogue \citep{james08}. The ${T_\textrm{gas}}$ are quoted from ${T_\textrm{kin}}$ in \citet{cesaroni92}.}
\label{uchiisourceinfo}
\end{deluxetable}

\begin{deluxetable}{lccccccccc}
\tablewidth{0pt}
\tabletypesize{\tiny}
\tablecolumns{          10}
\tablecaption{qIRDCc Observed Line Parameters}
\scriptsize
\tablehead{\colhead{Source Name} & \colhead{Molecule} & \colhead{Transition} & \colhead{${T_\textrm{MB}}$} & \colhead{$T_\textrm{rms}$} & \colhead{$\bigtriangleup$v} & \colhead{$\int {T_\textrm{MB}}$dv} & \colhead{$v_\textrm{LSR}$} & \colhead{$\tau_\textrm{thin}$} & \colhead{$T_\textrm{ex}$}\\
\colhead{}  &	\colhead{}  &	\colhead{} & \colhead{[K] }      & \colhead{[K] }      & \colhead{[km$s^{-1}$]} & \colhead{[K km$s^{-1}$]} & \colhead{[km$s^{-1}$]} & \colhead{} & \colhead{[K]}}
\startdata
         G018.82$\_$MM6      &          \hcn\	& $F=0-1$ &   0.039      &  0.034      &  2.407      (  1.710      )      &  0.099      (  0.054      )      & 65.7       &\nodata&\nodata \\[-1ex]
     	        				  &              & $F=2-1$ &   0.132      &  0.034      &  1.302      (  0.410      )      &  0.183      (  0.067      )      &&&\\[-1ex]
     	        				  &              & $F=1-1$ &   0.085      &  0.034      &  6.946      (  1.220      )      &  0.626      (  0.118      )      &&&\\[-1ex]
     	        & \hnc\  &	  &   0.544      &  0.057      &  1.751      (  0.139      )      &  1.014      (  0.073      )      &&\nodata&\nodata\\[1.5ex]
         G028.37$\_$MM9      &          \hcn\	& $F=0-1$ &   0.047      &  0.047      &  3.222      (  1.610      )      &  0.162      (  0.078      )      & 80.1       &0.13& 12.55 \\[-1ex]
     	        &              & $F=2-1$ &   0.190      &  0.047      &  3.267      (  0.487      )      &  0.662      (  0.088      )      &&&\\[-1ex]
     	        &              & $F=1-1$ &   0.141      &  0.047      &  3.457      (  0.691      )      &  0.518      (  0.091      )      &&&\\[-1ex]
     	        & \hnc\  &	  &   0.625      &  0.060      &  2.548      (  0.157      )      &  1.695      (  0.093      )      &      &0.14&13.39\\
\enddata
\label{qirdcdrvdlinepara}
\end{deluxetable}

\begin{deluxetable}{lccccccccc}
\tablewidth{0pt}
\tabletypesize{\tiny} 
\tablecolumns{          10}  
\tablecaption{AIRDC Observed Line Parameters}  
\scriptsize
\tablehead{\colhead{Source Name} & \colhead{Molecule} & \colhead{Transition} & \colhead{${T_\textrm{MB}}$} & \colhead{$T_\textrm{rms}$} & \colhead{$\bigtriangleup$v} & \colhead{$\int {T_\textrm{MB}}$dv} & \colhead{$v_\textrm{LSR}$}& \colhead{$\tau_\textrm{thin}$} & \colhead{$T_\textrm{ex}$} \\
\colhead{}  &	\colhead{}  &	\colhead{} & \colhead{[K] }      & \colhead{[K] }      & \colhead{[km$s^{-1}$]} & \colhead{[K km$s^{-1}$]} & \colhead{[km$s^{-1}$]}& \colhead{}	&	\colhead{[K]}}
\startdata
         G019.27$\_$MM2      &          \hcn\	& $F=0-1$ &   0.104      &  0.052      &  1.881      (  0.746      )      &  0.209      (  0.072      )      & 26.7      &          0.16      &         10.22       \\[-1ex]
     	        &              & $F=2-1$ &   0.263      &  0.052      &  2.388      (  0.393      )      &  0.668      (  0.091      )      &      &                     &                      \\[-1ex]
     	        &              & $F=1-1$ &   0.231      &  0.052      &  2.315      (  0.722      )      &  0.570      (  0.115      )      &      &                     &                      \\[-1ex]
     	        & \hnc\  &	  &   0.497      &  0.069      &  1.578      (  0.214      )      &  0.834      (  0.090      )      &      &          0.18      &          7.00      \\[1.5ex]
         G022.35$\_$MM1      &          \hcn\	& $F=0-1$ &   0.145      &  0.063      &  1.488      (  0.439      )      &  0.229      (  0.071      )      & 52.6      &          0.16      &         19.70       \\[-1ex]
     	        &              & $F=2-1$ &   0.311      &  0.063      &  2.988      (  0.427      )      &  0.990      (  0.115      )      &      &                     &                      \\[-1ex]
     	        &              & $F=1-1$ &   0.203      &  0.063      &  3.492      (  0.789      )      &  0.756      (  0.130      )      &      &                     &                      \\[-1ex]
     	        & \hnc\  &	  &   0.391      &  0.049      &  2.191      (  0.248      )      &  0.913      (  0.078      )      &      &        \nodata      &        \nodata      \\
\enddata
\label{airdcdrvdlinepara}
\end{deluxetable}

\begin{deluxetable}{lccccccccccc}
\tablewidth{0pt}
\tabletypesize{\tiny}
\tablecolumns{          12}  
\tablecaption{HMPO Observed Line Parameters}  
\scriptsize
\tablehead{\colhead{Source Name} & \colhead{Molecule} & \colhead{Transition} & \colhead{${T_\textrm{MB}}$} & \colhead{$T_\textrm{rms}$} & \colhead{$\bigtriangleup$v} & \colhead{$\int {T_\textrm{MB}}$dv} & \colhead{$v_\textrm{LSR}$}& \colhead{$\tau_\textrm{thin}$} & \colhead{$T_\textrm{ex}$} \\
\colhead{}  &	\colhead{}  &	\colhead{} & \colhead{[K] }      & \colhead{[K] }      & \colhead{[km$s^{-1}$]} & \colhead{[K km$s^{-1}$]} & \colhead{[km$s^{-1}$]}& \colhead{}	&	\colhead{[K]}}
\startdata
        IRAS05358+3543      &         \hcn\	& $F=0-1$ &   0.075      &  0.044      &  6.992      (  2.300      )      &  0.557      (  0.166      )      &-17.4      &          0.16      &         23.50       \\[-1ex]
     	        &              & $F=2-1$ &   0.444      &  0.044      &  2.001      (  0.202      )      &  0.946      (  0.121      )      &      &                     &                      \\[-1ex]
     	        &              & $F=1-1$ &   0.188      &  0.044      &  3.823      (  0.476      )      &  0.767      (  0.087      )      &      &                     &                      \\[-1ex]
     	        & \hnc\  &	  &   0.479      &  0.065      &  1.755      (  0.213      )      &  0.895      (  0.088      )      &      &          0.02      &         19.74      \\[1.5ex]
        IRAS05373+2349      &          \hcn\	& $F=0-1$ &   0.199      &  0.068      &  0.776      (  0.620      )      &  0.165      (  0.079      )      &  2.4      &          0.07      &         26.27       \\[-1ex]
     	        &              & $F=2-1$ &   0.479      &  0.068      &  1.901      (  0.356      )      &  0.969      (  0.118      )      &      &                     &                      \\[-1ex]
     	        &              & $F=1-1$ &   0.259      &  0.068      &  1.240      (  0.409      )      &  0.342      (  0.083      )      &      &                     &                      \\[-1ex]
     	        & \hnc\  &	  &   0.485      &  0.059      &  1.638      (  0.221      )      &  0.845      (  0.083      )      &      &          0.07      &         15.24      \\
\enddata
\tablecomments{$^{a}$The detection of the \hnc\ line can be confirmed by eye even though their SNR is smaller than 3.  }
\label{hmpodrvdlinepara}
\end{deluxetable}

\begin{deluxetable}{lccccccccccc}
\tablewidth{0pt}
\tabletypesize{\tiny}
\tablecolumns{          12}  
\tablecaption{UCHII Observed Line Parameters}  
\scriptsize
\tablehead{\colhead{Source Name} & \colhead{Molecule} & \colhead{Transition} & \colhead{${T_\textrm{MB}}$} & \colhead{$T_\textrm{rms}$} & \colhead{$\bigtriangleup$v} & \colhead{$\int {T_\textrm{MB}}$dv} & \colhead{$v_\textrm{LSR}$}& \colhead{$\tau_\textrm{thin}$} & \colhead{$T_\textrm{ex}$} \\
\colhead{}  &	\colhead{}  &	\colhead{} & \colhead{[K] }      & \colhead{[K] }      & \colhead{[km$s^{-1}$]} & \colhead{[K km$s^{-1}$]} & \colhead{[km$s^{-1}$]}& \colhead{}	&	\colhead{[K]}}
\startdata
           IRAS02232+6138      &         \hcn\	& $F=0-1$ &   0.288      &  0.042      &  3.422      (  0.361      )      &  1.049      (  0.083      )      &-46.4      &          0.15      &         59.33       \\[-1ex]
     	        &              & $F=2-1$ &   1.160      &  0.042      &  3.454      (  0.093      )      &  4.269      (  0.096      )      &      &                     &                      \\[-1ex]
     	        &              & $F=1-1$ &   0.658      &  0.042      &  3.381      (  0.164      )      &  2.369      (  0.095      )      &      &                     &                      \\[-1ex]
     	        & \hnc\  &	  &   0.378      &  0.039      &  2.938      (  0.185      )      &  1.182      (  0.065      )      &      &          0.05      &         29.14      \\[1.5ex]
           IRAS02575+6017      &          \hcn\	& $F=0-1$ &   0.096      &  0.048      &  2.062      (  0.662      )      &  0.211      (  0.065      )      &-38.0      &          0.04      &         47.81       \\[-1ex]
     	        &              & $F=2-1$ &   0.543      &  0.048      &  2.347      (  0.154      )      &  1.356      (  0.074      )      &      &                     &                      \\[-1ex]
     	        &              & $F=1-1$ &   0.228      &  0.048      &  2.396      (  0.392      )      &  0.583      (  0.076      )      &      &                     &                      \\[-1ex]
     	        & \hnc\  &	  &   0.300      &  0.041      &  1.381      (  0.154      )      &  0.441      (  0.046      )      &      &          0.01      &         15.98      \\
\enddata
\tablecomments{$^{a}$The detection of the \hnc\ line can be confirmed by eye even though their SNR is smaller than 3.}
\label{uchiidrvdlinepara}
\end{deluxetable}

\begin{deluxetable}{crrrr}
\tablewidth{0pt}
\tabletypesize{\footnotesize}
\tablecolumns{        5  } 
\tablecaption{Derived mean values \textbf{and standard deviations} of $\tau_\textrm{thin}$ and  $ T_\textrm{ex} $ for each evolutionary phases. }
\scriptsize
\tablehead{\colhead{Evolutionary Phases} & \colhead{$\tau_\textrm{ \hcn\ }$} & \colhead{$T_\textrm{ex}$(HCN)} & \colhead{$\tau_\textrm{ \hnc\ }$} &\colhead{$T_\textrm{ex}$(HNC)}  \\
\colhead{}  & & \colhead{[K]} & & \colhead{[K]}}
\startdata
qIRDCc & 0.11~(0.03) &18.79~(8.74) & 0.11~(0.03) & 15.10~(2.58)\\[-1ex]
aIRDCc & 0.16~(0.08) & 22.38~(13.33) & 0.13~(0.08) & 15.77~(7.03)\\[-1ex]
HMPO &  0.09~(0.05) & 30.88~(15.77) & 0.08~(0.05) & 16.99~(6.11)\\[-1ex]
UCHII & 0.18~(0.29) & 43.92~(16.12)& 0.07~(0.06) & 24.67~(15.26) \\[-1ex]
\enddata
\label{meanTauandTex}
\end{deluxetable}

\begin{deluxetable}{l r r r r r r }
\tablewidth{0pt}
\tabletypesize{\tiny}
\tablecolumns{          7}  
\tablecaption{QIRDC Derived Line Parameters}  
\scriptsize
\tablehead{\colhead{Source Name} & \colhead{N(HCN)} & \colhead{N(HNC)} & \colhead{$\frac{\textrm{HCN}}{\textrm{HNC}}$} & \colhead{N(\IRASolec)} & \colhead{X(HCN)} & \colhead{X(HNC)} \\
\colhead{} & \colhead{[$10^{14}cm^{-2}$]} & \colhead{[$10^{14}cm^{-2}$]} & \colhead{} & \colhead{[$10^{22}cm^{-2}$]} & \colhead{[$10^{-8}$]} & \colhead{[$10^{-8}$]}}
\startdata
                     G018.82$\_$MM6    &    1.1    (    0.11    )    &    1.1    (    0.09    )    &   1.06    (   0.156    )    &             \nodata    &             \nodata    &             \nodata            \\[-1ex]
                     G028.37$\_$MM9    &    1.2    (    0.07    )    &    1.4    (    0.05    )    &   0.87    (   0.069    )    &   1.75    &    0.7    &    0.8            \\[-1ex]
                    G028.37$\_$MM17    &    1.5    (    0.12    )    &    1.5    (    0.07    )    &   1.02    (   0.141    )    &   0.68    &    2.2    &    2.1            \\[-1ex]
                     G028.53$\_$MM7    &    0.8    (    0.13    )    &    1.0    (    0.10    )    &   0.81    (   0.130    )    &             \nodata    &             \nodata    &             \nodata            \\[-1ex]
                     G031.97$\_$MM9    &    2.0    (    0.07    )    &    1.5    (    0.05    )    &   1.30    (   0.116    )    &   1.27    &    1.5    &    1.2            \\[-1ex]
                     G033.69$\_$MM8    &    1.0    (    0.12    )    &    0.9    (    0.07    )    &   1.17    (   0.162    )    &   0.55    &    1.9    &    1.6            \\[-1ex]
                     G034.77$\_$MM4    &    1.0    (    0.16    )    &    1.3    (    0.06    )    &   0.77    (   0.132    )    &             \nodata    &             \nodata    &             \nodata            \\[-1ex]
                     G035.39$\_$MM5    &    1.1    (    0.10    )    &    1.5    (    0.06    )    &   0.75    (   0.087    )    &             \nodata    &             \nodata    &             \nodata            \\
\enddata
\label{qirdccoldentable}
\end{deluxetable}

\begin{deluxetable}{l r r r r r r }
\tablewidth{0pt}
\tabletypesize{\tiny}
\tablecolumns{          7}  
\tablecaption{AIRDC Derived Line Parameters}  
\scriptsize
\tablehead{\colhead{Source Name} & \colhead{N(HCN)} & \colhead{N(HNC)} & \colhead{$\frac{\textrm{HCN}}{\textrm{HNC}}$} & \colhead{N(\IRASolec)} & \colhead{X(HCN)} & \colhead{X(HNC)} \\
\colhead{} & \colhead{[$10^{14}cm^{-2}$]} & \colhead{[$10^{14}cm^{-2}$]} & \colhead{} & \colhead{[$10^{22}cm^{-2}$]} & \colhead{[$10^{-8}$]} & \colhead{[$10^{-8}$]}}
\startdata
                     G019.27$\_$MM2    &    1.3    (    0.08    )    &    0.7    (    0.08    )    &   1.90    (   0.210    )    &   0.59    &    2.2    &    1.1            \\[-1ex]
                     G022.35$\_$MM1    &    2.7    (    0.06    )    &    1.0    (    0.07    )    &   2.75    (   0.245    )    &             \nodata    &             \nodata    &             \nodata            \\[-1ex]
                     G023.60$\_$MM1    &    6.8    (    0.03    )    &    2.1    (    0.03    )    &   3.24    (   0.127    )    &             \nodata    &             \nodata    &             \nodata            \\[-1ex]
                     G024.60$\_$MM1    &    2.8    (    0.06    )    &    0.7    (    0.10    )    &   4.20    (   0.492    )    &   1.08    &    2.6    &    0.6            \\[-1ex]
                     G028.37$\_$MM4    &    3.8    (    0.04    )    &    2.9    (    0.04    )    &   1.31    (   0.065    )    &   1.53    &    2.5    &    1.9            \\[-1ex]
                     G028.37$\_$MM6    &    2.5    (    0.04    )    &    1.2    (    0.04   )    &   2.01    (   0.115    )    &   1.13    &    2.2    &    1.1            \\[-1ex]
                     G030.97$\_$MM1    &    3.2    (    0.05    )    &    1.3    (    0.05    )    &   2.34    (   0.173    )    &   1.04    &    3.0    &    1.3            \\[-1ex]
                     G031.97$\_$MM1    &   23.1    (    0.01    )    &    3.6    (    0.03    )    &   6.45    (   0.187    )    &   3.51    &    6.6    &    1.0            \\
\enddata
\label{airdccoldentable}
\end{deluxetable}

\begin{deluxetable}{l  r r r r r r r}
\tablewidth{0pt}
\tabletypesize{\tiny}
\tablecolumns{          8}  
\tablecaption{HMPO Derived Line Parameters}  
\scriptsize
\tablehead{\colhead{Source Name} & \colhead{N(HCN)} & \colhead{N(HNC)} & \colhead{$\frac{\textrm{HCN}}{\textrm{HNC}}$} & \colhead{N(\IRASolec)} & \colhead{X(HCN)} & \colhead{X(HNC)} \\
\colhead{} & \colhead{[$10^{14}cm^{-2}$]} & \colhead{[$10^{14}cm^{-2}$]} & \colhead{} & \colhead{[$10^{22}cm^{-2}$]} & \colhead{[$10^{-8}$]} & \colhead{[$10^{-8}$]}}
\startdata
                     IRAS05358+3543    &   11.2    (    0.04    )    &    3.5    (    0.11    )    &   3.22    (   0.376    )    &   2.54    &    4.4    &    1.4            \\[-1ex]
                     IRAS05373+2349    &    4.5    (    0.07    )    &    1.8    (    0.09    )    &   2.44    (   0.276    )    &   0.85    &    5.3    &    2.2            \\[-1ex]
                     IRAS18024-2119    &    5.6    (    0.04    )    &    1.9    (    0.05    )    &   2.94    (   0.186    )    &   1.44    &    3.9    &    1.3            \\[-1ex]
                     IRAS18089-1732    &   13.4    (    0.02    )    &    3.2    (    0.04    )    &   4.25    (   0.193    )    &   3.29    &    4.1    &    1.0            \\[-1ex]
                     IRAS18102-1800    &    3.1    (    0.05    )    &    1.1    (    0.06    )    &   2.95    (   0.234    )    &   2.00    &    1.6    &    0.5            \\[-1ex]
                     IRAS18144-1723    &   10.9    (    0.04    )    &    3.2    (    0.04    )    &   3.44    (   0.186    )    &   1.13    &    9.7    &    2.8            \\[-1ex]
                     IRAS18151-1208    &    6.9    (    0.08    )    &    1.3    (    0.07    )    &   5.21    (   0.536    )    &   1.67    &    4.2    &    0.8            \\[-1ex]
                     IRAS18182-1433    &   12.2    (    0.03    )    &    1.4    (    0.06    )    &   8.98    (   0.604    )    &   2.00    &    6.1    &    0.7            \\
\enddata
\label{hmpocoldentable}
\end{deluxetable}

\begin{deluxetable}{l  r r r r r r }
\tablewidth{0pt}
\tabletypesize{\tiny}
\tablecolumns{          7}  
\tablecaption{UCHII Derived Line Parameters}  
\scriptsize
\tablehead{\colhead{Source Name} & \colhead{N(HCN)} & \colhead{N(HNC)} & \colhead{$\frac{\textrm{HCN}}{\textrm{HNC}}$} & \colhead{N(\IRASolec)} & \colhead{X(HCN)} & \colhead{X(HNC)} \\
\colhead{} & \colhead{[$10^{14}cm^{-2}$]} & \colhead{[$10^{14}cm^{-2}$]} & \colhead{} & \colhead{[$10^{22}cm^{-2}$]} & \colhead{[$10^{-8}$]} & \colhead{[$10^{-8}$]}}
\startdata
                     IRAS02232+6138    &   48.3    (    0.01    )    &    3.7    (    0.05    )    &  12.99    (   0.650    )    &   3.88    &   12.4    &    1.0            \\[-1ex]
                     IRAS02575+6017    &   10.8    (    0.04    )    &    0.8    (    0.13    )    &  13.83    (   1.803    )    &   0.57    &   19.0    &    1.4            \\[-1ex]
                     IRAS03035+5819    &    4.5    (    0.04    )    &    0.2    (    0.19    )    &  18.08    (   3.542    )    &   0.50    &    8.8    &    0.5            \\[-1ex]
                     IRAS06053-0622    &    6.3    (    0.05    )    &    0.4    (    0.16    )    &  15.35    (   2.573    )    &             \nodata    &             \nodata    &             \nodata            \\[-1ex]
                     IRAS06084-0611    &   33.0    (    0.01    )    &    2.8    (    0.03    )    &  11.88    (   0.435    )    &   1.80    &   18.3    &    1.5            \\[-1ex]
                     IRAS06056+2131    &   11.5    (    0.04    )    &    2.0    (    0.06    )    &   5.86    (   0.386    )    &   1.53    &    7.6    &    1.3            \\[-1ex]
                     IRAS06058+2138    &   16.8    (    0.03    )    &    2.0    (    0.06    )    &   8.35    (   0.561    )    &   1.49    &   11.3    &    1.4            \\[-1ex]
                     IRAS06061+2151    &    4.6    (    0.04    )    &    0.2    (    0.17    )    &  21.27    (   3.686    )    &   0.48    &    9.7    &    0.5            \\
\enddata
\label{uchiicoldentable}
\end{deluxetable}

\begin{deluxetable}{llc}
\tablewidth{0pt}
\tabletypesize{\footnotesize}
\tablecolumns{          3} 
\tablecaption{Mean abundance ratio for each phases.} 
\scriptsize
\tablehead{\colhead{Evolutionary Phases} && \colhead{Mean \ratio}}
\startdata
IRDC & Quiescent core & 0.97($\pm0.10$) \\[-1ex]
 &Active core& 2.65($\pm0.88$) \\[-1ex]
HMPO& & 4.17($\pm1.03$) \\[-1ex]
UCHII& & 8.96($\pm3.32$) \\[-1ex]
\enddata
\label{drvdratio}
\end{deluxetable}

\begin{deluxetable}{llccc}
\tablewidth{0pt}
\tabletypesize{\footnotesize}
\tablecolumns{        5  } 
\tablecaption{Conditions for chemical modeling}
\scriptsize
\tablehead{\colhead{Evolutionary Phases} &\colhead{$n$} & \colhead{$T_\textrm{D}$} & \colhead{$T_\textrm{gas}$} & Chemical Timescale \\
\colhead{} & \colhead{[$cm^{-3}$]} & \colhead{[K]} & \colhead{[K]} & \colhead{[yr]}}
\startdata
IRDC & 			& 27 & 27 & 10000 \\[-1ex]
HMPO& 10$^{5} $    & 50 & 50 & 60000\\[-1ex]
UCHII& 			& 100 & 100 & 50000\\[-1ex]
\enddata
\label{modelcondition}
\end{deluxetable}


\begin{thebibliography}{}

\bibitem[Aalto et al.(2012)]{aalto12}
Aalto, S., Garcia-Burillo, S., Winters, J.M., et al.
2012, \aap, 537, A44

\bibitem[Bechtel et al.(2006)]{bechtel06} 
Bechtel, H.~A., Steeves, A.~H., \& Field, R.~W.\ 2006, \apjl, 649, L53 

\bibitem[Bergin et al.(1997)]{bergin97}
Bergin, E.A., Ungerechts, H., Goldsmith, P.F., et al.
1997, \apj, 482, 267

\bibitem[Bergin \& Tafalla(2007)]{berginandtafalla07} 
Bergin, E.~A., \& Tafalla, M.\ 2007, \araa, 45, 339 

\bibitem[Beuther et al.(2002)]{beuther02}
Beuther, H., Schilke, P., Menten, K.M., et al.
2002, \apjs, 566, 945

\bibitem[Bhattacharya \& Gordy(1960)]{bhattacharyaandgordy60}
Bhattacharya, B. N., \& Gordy, W.
1960, Phys. Rev., 119, 144

\bibitem[Blackman et al.(1976)]{blackman76}
Blackman, G. L., Brown, R. D., Godfrey, P. D., \& Gunn, H. I.
1976, \nat, 261, 395

\bibitem[Cazaux \aand\ Tielens(2004)]{cazauxandtielens04}
Cazaux, S., \aand\ Tielens, A.G.G.M.
2004, \apj, 604, 222

\bibitem[Cazaux \aand\ Tielens(2010)]{cazauxandtielens10}
Cazaux, S., \aand\ Tielens, A.G.G.M.
2010, \apj, 715, 698

\bibitem[Cernicharo et al.(1984)]{cernicharo84}
Cernicharo, J., Castets, A., Duvert, G., \aand Guilloteau, S.
1984, \aap, 139, L13

\bibitem[Cernicharo \aand Guelin(1987)]{cernicharoandguelin87}
Cernicharo, J., \aand Guillin, M.
1987, \aap, 183, 10

\bibitem[Cesaroni et al.(1992)]{cesaroni92}
Cesaroni, R., Walmsley, C.M., \&\ Churchwell, E.
1992, \apjs, 256, 618

\bibitem[Chambers et al.(2009)]{chambers09}
Chambers, E.T., Jackson, J.M., Rathborne, J.M., \aand Simon, R. 
2009, \apjs, 181, 360

\bibitem[Chira et al.(2013)]{chira13}
Chira, R.-A., Beuter, H., Linz, H., et al.
2013, \aap, 552, AA40

\bibitem[Churchwell et al.(1990)]{churchwell90}
Churchwell, E., Walmsley, C.~M., \& Cesaroni, R.
1990, \aaps, 83, 119

\bibitem[Churchwell et al.(1992)]{churchwell92} 
Churchwell, E., Walmsley, C.~M., \& Wood, D.~O.~S.
1992, \aap, 253, 541

\bibitem[Churchwell, Ed.(2002)]{churchwell02}
Churchwell, Ed.
2002, \araa, 40, 27

\bibitem[Di Francesco et al.(2008)]{james08}
Di Francesco, J., Johnstone, D., Kirk, H. M., MacKenzie, T., \aand Ledwosinska, E. 
2008, \apjs, 175,277

\bibitem[Egan et al.(1998)]{egan98}
Egan, M.P., Shimpman, R.F., Price, S.D., Carey, S.J., \aand Clark, F.O.
1998, \apj, 494, L199

\bibitem[Garcia-Segura et al.(1996)]{segura96} 
Garcia-Segura, G., \aand Franco, J.
1996, \apj, 469, 171

\bibitem[Gerner et al.(2014)]{gerner14}
Gerner, T., Beuther, H., Semenov, D., et al.
2014, \aap, 563A, 97G

\bibitem[Goldsmith et al.(1986)]{goldsmith86} 
Goldsmith, P.F., Irvine, W.M., Hjalmarson, A., \aand Ellder, J.
1986, \apj, 705,123 

\bibitem[Graninger et al.(2014)]{graninger14} 
Graninger, D.M., Herbst, E., Oberg, K.I., \aand Vasyunin, A.I.
2014, \apj, 787,74G

\bibitem[Harada et al.(2010)]{harada10}
Harada, N., Herbst, E., \aand Wakelam, V.
2010, \apj, 721, 1570

\bibitem[Harada et al.(2012)]{harada12}
Harada, N., Herbst, E., \aand Wakelam, V.
2012, \apj, 756, 104

\bibitem[Harjunp{\"a}{\"a} et al.(2004)]{harjunpaa04} 
Harjunp{\"a}{\"a}, P., Lehtinen, K., \& Haikala, L.~K.\ 
2004, \aap, 421, 1087 

\bibitem[Hirota et al.(1998)]{hirota98} 
Hirota, T., Yamamoto, S.,Mikami, H., \& Ohishi, M. 
1998, \apj, 503, 717

\bibitem[Hoq et al.(2013)]{hoq13}
Hoq, S., Jackson, J.M., Foster, J.B., et al.
2013, \apj, 777, 157

\bibitem[Kauffmann et al.(2008)]{kauffmann08} 
Kauffmann, J., Bertoldi, F., Bourke, T.L., Evans, N.J., \aand Lee, C.H. 
2008, \aap, 487, 993

\bibitem[Kim \aand Koo(2001)]{ktkim01} 
Kim, K.-T., \aand Koo, B.-C.
2001, \apj, 549, 979

\bibitem[Kim et al.(2011)]{ktkim11} 
Kim, K.-T., Byun, D.-Y., Je, D.-H., et al. 
2011, J. Korean Astron. Soc., 44, 81

\bibitem[Kurtz et al.(1994)]{kurtz94}
Kurtz, S., Churchwell, E., \aand Wood, D.O.S. 
1994, \apjs, 91, 659

\bibitem[Lee et al.(2003)]{lee03}
Lee, J.-E., Evans, N.J., \aand Shirley, Y.L. 
2003, \apj, 583, 789

\bibitem[Lee et al.(2004)]{lee04}
Lee, J.-E., Edwin, A.B., \aand Evans, N.J.
2004, \apj, 617, 360

\bibitem[Lee et al.(2011)]{lee11}
Lee, S.-S., Byun, D.-Y., Oh, C. S., et al.
2011, PASP, 123, 1398

\bibitem[Loughnane et al.(2012)]{loughnane12}
Loughnane, R.M., Redman, M.P., Thompson, M.A., et al.
2012, \mnras, 420, 1367 

\bibitem[Mangum \aand Shirley(2015)]{mangumandshirley15}
Mangum, J.~G., \& Shirley, Y.~L.\ 2015, arXiv:1501.01703 

\bibitem[McElroy et al.(2013)]{mcelroy13}
McElroy, D., Walsh, C., Markwick, A.J., et al.
2013, \aap, 550A, 36M

\bibitem[Mendes et al.(2012)]{mendes12}
Mendes, M.B., Buhr, H., Berg, M.H., et al.
2012, \apj, 746L, 8M

\bibitem[Miettinen et al.(2014)]{miettinen14}
Miettinen, O.
2014, \aap, 562, A3

\bibitem[Molinari et al.(1996)]{molinari96}
Molinari, S., Brand, J., Cesaroni, R., \aand Palla, F.
1996, \aap, 308, 573

\bibitem[Molinari et al.(2000)]{molinari00}
Molinari, S., Brand, J., Cesaroni, R., \aand Palla, F.
2000, \aap, 355, 617

\bibitem[Peretto \aand Fuller(2009)]{perettoandfuller09}
Peretto, N., \aand Fuller, G.A.
2009, \aap, 505, 405

\bibitem[Purcell et al.(2006)]{purcell06}
Purcell, C. R., Balasubramananyam, R., Burton, M. G., et al. 
2006, \mnras, 376, 553

\bibitem[Ragan et al.(2011)]{ragan11}
Ragan, S.E., Bergin, E.A., \& Wilner, D.
2011, \apj, 736, 163

\bibitem[Ramesh \aand\ Sridharan(1997)]{rameshandsridharan97}
Ramesh, B. \aand\ Sridharan, T.K.
1997, \mnras, 284, 1001

\bibitem[Rathborne et al.(2006)]{rathborne06}
Rathborne, J.M., Jackson, J.M., \aand Simon, R.
2006, \apj, 641, 389 

\bibitem[Rathborne et al.(2010)]{rathborne10}
Rathborne, J.M., Jackson, J.M., Chambers, E.T., et al.
2010, \apj, 715, 310

\bibitem[Rathborne et al.(2014)]{rathborne14}
Rathborne, J.M., Longmore, S.N., Jackson, J.M., et al.
2014, \apj, 786, 140

\bibitem[Sakai et al.(2008)]{sakai08}
Sakai, T., Sakai, N., Kamegai, K., \aand Hirota, T.
2008, \apj, 678, 1049

\bibitem[Sakai et al.(2010)]{sakai10}
Sakai, T., Sakai, N., Hirota, T., \aand Yamamoto, S.
2010, \apj, 714, 1658

\bibitem[Sanhueza et al.(2012)]{sanhueza12}
Sanhueza, P., Jackson, J.M., Foster, J.B., et al.
2012, \apj, 755, 60

\bibitem[Schilke et al.(1992)]{schilke92}
Schilke, P., Walmsley, C.M., Pineau Des Forets, G., et al.
1992, \aap, 256, 595

\bibitem[Semenov et al.(2010)]{semenov10}
Semenov, D., Hersant, F., Wakelam, V., et al. 
2010, \aap, 552, A42

\bibitem[Simon et al.(2006a)]{simon06a}
Simon, R., Jackson, J.M., Rathborne, J.M., \aand Chambers, E.T. 
2006a, \apj, 639, 227

\bibitem[Simon et al.(2006b)]{simon06b}
Simon, R., Rathborne, J.M., Shah, R.Y., Jackson, J.M., \aand Chambers, E.T. 
2006b, \apj, 653, 1325

\bibitem[Sridharan et al.(2002)]{sridharan02}
Sridharan, T.K., Beuther, H., Schilke, P., Menten, K.M., \aand Wyrowski, F 
2002, \apj, 566, 931

\bibitem[Sohn et al.(2007)]{sohn07}
Sohn, J., Lee, C. W., Park, Y.S., et al.
2007, \apj, 664, 928

\bibitem[Talbi et al.(1996)]{talbi96}
Talbi, D., Ellinger, Y., \aand Herbst, E.
1996, \aap, 314, 688

\bibitem[Thompson et al.(2006)]{thompson06} 
Thompson, M.A., Hatchell, J., Walsh, A.J., Macdonald, G.H., \aand Millar, T.J.
2006, \aap, 453, 1003

\bibitem[van der Tak et al.(2009)]{vandertak09} 
van der Tak, F.~F.~S., M{\"u}ller, H.~S.~P., Harding, M.~E., \& Gauss, J.
2009, \aap, 507, 347

\bibitem[Vasyunina et al.(2011)]{vasyunina11}
Vasyunina, T., Linz, H., Henning, T., et al. 
2011, \aap, 527, A88

\bibitem[Wilson \aand Rood(1994)]{wilsonrood94}
Wilson, T.L., \aand Rood, R.T. 
1994, \araa, 32, 191

\bibitem[Wood \aand Churchwell(1989)]{woodchurchwell89}
Wood, D.O.S., \aand Churchwell, E. 
1989, \apjs, 69, 831

\bibitem[Wu \aand Evans(2003)]{wuandevans03}
Wu, J. \aand Evans, N.J.
2003, \apj, 592, L79

\bibitem[Zhang et al.(2005)]{zhang05}
Zhang, Q., Hunter, T.R., Brand, J., et al.
2005, \apj, 625, 864

\bibitem[Zinnecker \aand York (2007)]{zinnecker07}
Zinnecker, H. \aand York, H. W. 
2007, \araa, 45, 481 


\end{thebibliography}
\end{document}